\documentclass[useAMS,usenatbib]{mn2e}
\usepackage{amsmath}
\usepackage{amssymb}

\usepackage{graphics}
\usepackage{graphicx}
\usepackage{epstopdf}
\usepackage{footnote}
\usepackage{tablefootnote}
\usepackage{txfonts}
\usepackage{lipsum}
\usepackage{float}
\usepackage[draft]{hyperref}
\usepackage{macros}
\usepackage{fixltx2e}

\makesavenoteenv{tabular}

\DeclareSymbolFont{cmletters}{OML}{cmm}{m}{it}
\DeclareMathSymbol{v}{\mathalpha}{cmletters}{"76}

\usepackage[usenames,dvipsnames,svgnames,table]{xcolor}
\usepackage{hyperref}
\definecolor{darkblue}{rgb}{0.0,0.0,0.3}
\hypersetup{colorlinks,breaklinks,
            linkcolor=darkblue,urlcolor=darkblue,
            anchorcolor=darkblue,citecolor=darkblue}

\title[MRI Decretion Disc]{3D MHD Simulation of a Pulsationally-Driven MRI Decretion Disc
 }
\author[S. M. Ressler ]{S. M. Ressler$^{1}$\\
$^{1}$Kavli Institute for Theoretical Physics, University of California Santa Barbara, Santa Barbara, CA 93107 }

\begin{document}

\maketitle

\begin{abstract}
We explore the pulsationally driven orbital mass ejection mechanism for Be star disc formation using isothermal, 3D magnetohydrodynamic (MHD) and hydrodynamic simulations.
Non-radial pulsations are added to a star rotating at 95\% of critical as an inner boundary condition that feeds gas into the domain. 
In MHD, the initial magnetic field within the star is weak. 
The hydrodynamics simulation has limited angular momentum transport, resulting in repeating cycles of mass accumulation into a rotationally-supported disc at small radii followed by fall-back onto the star.
The MHD simulation, conversely, has efficient (Maxwell $\alpha_{\rm M}$ $\sim$ 0.04) angular momentum transport provided by both turbulent and coherent magnetic fields: a slowly decreting midplane driven by the magnetorotational instability and a supersonic wind on the surface of the disc driven by global magnetic torques.  
The angle and time-averaged properties near the midplane agree reasonably well with a 1D viscous decretion disc model with a modified $\tilde\alpha=0.5$, in which the gas transitions from a subsonic thin disc to a supersonic spherical wind at the critical point.  
1D models, however, cannot capture the multi-phase decretion/angular structure seen in our simulations.   
Our results demonstrate that, at least under certain conditions, non-radial pulsations on the surface of a rapidly rotating, weakly magnetized star can drive a Keplerian disc with the basic properties of the viscous decretion disc paradigm, albeit coupled to a laminar wind away from the midplane.
Future modeling of Be star discs should consider the possible existence of such a surface wind. 
   \end{abstract}

\section{Introduction}
Be stars are characterized by two distinct observational features that set them apart from other B-type stars: the presence of spectral emission lines (which are often double-peaked) and an excess of infrared continuum emission (see \citealt{Porter2003,Rivinius2013} for detailed reviews).  The widely accepted view is that these features are caused by the reprocessing of stellar light by a dense, Keplerian disc in the midplane.  Since Be stars are also found to be rotating at a significant fraction of break-up, these discs are likely fed by some mass ejection mechanism in the photosphere.  Some form of viscosity then transfers angular momentum outwards from the star, driving ``decretion'' (or ``excretion'').  

The leading model for the decretion process is the viscous decretion disc  \citep{Lee1991,Okazaki2001,Krticka2011}, which adapts the \citet{Shakura1973} prescription for viscosity to a disc where the flow is moving outwards instead of inwards.  This model is essentially agnostic as to the source of viscosity as long as it can be parameterized by an $\alpha$-type relation.   Accretion disc theory has long assumed that the magnetorotational instability (MRI, \citealt{BalbusHawley}) provides a ubiquitous physical mechanism to amplify magnetic fields to the point that they can efficiently transport angular momentum outwards, so it is reasonable to assume that the same process could be at work in decretion discs as well.  Any Keplerian disc will be unstable to the MRI as long as there is a weak magnetic field present, where `weak' here means that the most unstable wavelength is small enough to fit inside the disc.  \citet{Krticka2015} showed that these conditions are naturally met in the viscous decretion disc model as long as the magnetic field in the star is sufficiently weak, which seems to be the case given the lack of any robust detection of magnetic fields in Be stars \citep{Rivinius2013}.

The process by which the Be star feeds matter to the disc is up for debate.  Several mechanisms have been proposed, including a radiatively driven, rapidly rotating wind compressed by ram pressure \citep{Bjorkman1993,Owocki1994,Owocki1996},  explosive flares on the stellar surface \citep{Kroll1997,Balona2003}, localized radiative forces near bright spots \citep{Cranmer1996}, and non-radial pulsations \citep{Baade1988,Rivinius2003,Cranmer2009Be}.  It has also been suggested that magnetic fields with foot-points on the rotating star can spin up an otherwise more spherical wind to the point that it forms a Keplerian disc \citep{Cassinelli2002,ud_Doula2002,Balona2003,Brown2004,Brown2008}. All of these models have their various strengths and weaknesses in terms of their feasibility and agreement with observables (see \citealt{Owocki2006} and Section 4 of \citealt{Rivinius2013} for a summary).  In this work we choose to focus on the non-radial pulsation mechanism.

Most, if not all Be stars are believed to pulsate given the consistent detection of multi-periodicity in their light curves by both ground and space telescopes (see Section 3.2 of \citealt{Rivinius2013} and references therein).  The properties of these pulsations differ from object to object and depend on the age of star, with periods ranging from about 0.5 to 2 days.  Many are well described by non-radial g-modes, particularly those with wave numbers $l=2$, $|m|= 2$ (e.g., \citealt{Rivinius2003,Neiner2020}). 
For some Be stars, the presence of pulsation is observed to be correlated with mass ejection events (e.g., \citealt{Rivinius1998,Gutierrez2010,Stefl2003,Goss2011,Neiner2020}), giving credence to the hypothesis that the pulsations cause or at least enhance disc formation.  

This hypothesis, however, is not without its problems.  The biggest being that typical Be stars have sound speeds that are only $\sim$ 5\% of the critical rotation speed at the equator, which means that the star would have to be rotating at $\sim$ 95 \% of the critical rotation rate in order for pulsations to add enough azimuthal velocity to eject the photospheric gas (assuming that pulsations are limited by the sound speed).  Some statistical analysis of the observations suggests that a significant number of Be stars are rotating much more sub-critically than this requirement, with a distribution centered around 60--70\% critical  \citep{Cranmer2005,Zorec2016}.  Conversely, other authors emphasize the difficulty in measuring the rotation rate precisely due to gravity darkening (\citealt{Townsend2004}; see \citealt{Rivinius2013} for a discussion), and in fact there are many Be stars that appear to be rotating very close to critical (e.g., \citealt{Delaa2011,Meilland2012}).  This means that the pulsationally driven orbital mass loss model should be relevant for at least a fraction of Be stars, with a possibility that said fraction could be large.

Many authors have applied isothermal smoothed-particle hydrodynamics (SPH) with $\alpha$-viscosities to simulations of Be decretion discs in the context of Be/Neutron star binaries \citep{Okazaki2002,Hayasaki2004,Hayasaki2005,Hayasaki2006,Martin2014,Brown2019}, where the disc feeding process is treated phenomenologically as a source of fluid particles with a given angular momentum; this allows them to be mostly agnostic to both the precise physical feeding mechanism and the source of viscosity.   Magnetohydrodynamics (MHD) simulations of isolated Be discs have mainly focused on line driven winds that are torqued and/or compressed by a global, stellar magnetic field (AKA the rigid magnetosphere or magnetically torqued disc model, \citealt{ud_Doula2002,ud_doula2006,ud_doula2008,ud_doula2009}).    These simulations were unable to produce stable Keplerian discs, instead finding dense midplanes characterized by mostly radial inflow and outflow.  
More recently, \citet{Kee2016} presented preliminary results from 2D, isothermal hydrodynamic simulations of a rapidly rotating, pulsating star.  They found that for pulsations in which density and azimuthal velocity perturbations are in phase, mass is consistently ejected from the star, forming a Keplerian disc that grows in both mass and radius in time.  By the end of their simulations it is still not clear whether the discs will continue to expand to larger radii or whether they will remain localized to $\sim$ a few stellar radii while increasing in density.   Magnetic fields are also likely a key ingredient for angular momentum transport.

In this work we seek to extend the methodology of \citet{Kee2016} to 3D, isothermal MHD simulations of non-radially pulsating, isolated Be stars.  The star is treated as near critically rotating gas with spherical harmonic perturbations to density and azimuthal velocity.  We consider the effects of an initially weak magnetic field contained within the star instead of the strong large scale fields required for the magnetically torqued disc model. We focus on one particular set of pulsation parameters, highlighting the angular momentum transport and comparing hydrodynamics to MHD simulations.

This paper is outlined as follows.  \S \ref{sec:methods} outlines the numerical model, \S \ref{sec:analytic} presents some preliminary analytic estimates of key quantities, \S \ref{sec:results} describes the simulation results, \S \ref{sec:lim} discusses the limitations of these simulations, and \S \ref{sec:conc} concludes.

\section{Methods}
\label{sec:methods}
We use {\tt Athena++} \citep{Athena++}, a conservative, grid-based code that solves the equations of ideal MHD in spherical coordinates.   We make use of the HLLE Riemann solver \citep{Einfeldt1988} and piecewise parabolic reconstruction \citep{Colella1984}.

We model the rotating ``star'' as an inner boundary condition at $r=R_\star$, where $R_\star$ is the stellar radius.  For simplicity, this inner boundary is spherical, although in reality the stellar surface should be oblate due to the stronger centrifugal force near the equator.   We adopt the pulsation prescription of \citet{Kee2016} for the density and azimuthal velocity within the boundary:
\begin{equation}
  \begin{aligned}
  \rho(\theta,\varphi) &= \rho_0 \exp\left\{-\frac{1}{2} \left[\frac{\Omega_\star R_\star}{c_{\rm s}}\cos(\theta)\right]^2\right\} \\ &\times 10^{\left[\log_{10}\left(\frac{\rho_{\rm max}}{\rho_0}\right)\sin\left(\frac{2 {\rm \pi} t}{P} +m\varphi\right)P_l^m\left(\cos\theta\right)\right] }  \\
  v_\varphi (\theta, \varphi) &= \Omega_\star R_\star \sin(\theta) + v_{\varphi,{\rm pert}}\sin\left(\frac{2 {\rm \pi} t}{P} +m\varphi + \varphi_0\right)P_l^m\left(\cos\theta\right),
  \end{aligned}
  \label{eq:rho}
\end{equation}
where $\rho_0$ is the stellar surface density, $\Omega_\star$ is the angular velocity of the star (which is assumed to rotate as a solid body), $c_{\rm s}$ is the sound speed, $\rho_{\rm max}$ is the maximum density of the pulsations, $t$ is time, $P$ is the period of the pulsations, $m$ and $l$ are the azimuthal and polar wave number of the pulsations, $P_l^m$ is the associated Legendre polynomial, $v_{\rm rot}$ is the unperturbed azimuthal velocity, $v_{\varphi,{\rm pert}}$ is the amplitude of the perturbation in $v_\varphi$, and $\varphi_0$ is the phase of the pulsations. Here we use $\varphi_0 = 0$ and $m=2$, meaning that the density and velocity perturbations are in phase and have negative (retrograde) phase velocity, as well as $l = m$, so that $P_{2}^{2} = \sin(\theta)^2$.  The density perturbations are treated logarithmically, i.e., $\log_{10}(\rho /\rho_0)$ is proportional to a sinusoid instead of $\rho-\rho_0$. This is done in order to better represent the effect that pulsations would have on the exponentially decreasing density with radius near the surface of the star.  In practice, however, the relatively small amplitude of the perturbations that we use makes the difference between linear and logarithmic variation almost negligible.  The exponential dependence on $\cos^2(\theta)$ in Equation \eqref{eq:rho} is such that the pressure and centrifugal forces are balanced in the $\theta$ direction.  This helps prevent the gas that is ejected from the boundary into the domain from immediately collapsing or expanding in that direction.\footnote{For example, if the density/pressure within the boundary were instead uniform, gas would enter the domain through the polar region of the inner boundary (not because of pulsations but because of the large pressure gradient with what is essentially floors), collapse towards the midplane, then tend to fall back onto the star and interfere with the pulsational feeding.  
}

We assume that the surface of the un-perturbed star is rotating at a constant angular velocity  near the surface that is close to break-up, characterized by $W = (v_{\rm rot}/v_{\rm kep})(r=R_\star,\theta={\rm \pi}/2) = 0.95$, where $v_{\rm kep} = \sqrt{GM_\star/r}$ is the Keplerian velocity, $M_\star$ is the mass of the star, and $G$ is the gravitational constant. We then set $v_{\rm rot} = W \sqrt{GM_\star/R_\star} \sin(\theta)$.   The amplitudes of the perturbations are set as $\rho_{\rm max} = 1.1 \rho_0$ and $v_{\varphi,{\rm pert}} = 0.055 v_{\rm kep}(R_\star)$, while the period is $P = 4.11 t_\star$, where $t_\star \equiv \sqrt{R_\star^3/(GM_\star)}$.  Inside the boundary the radial and polar velocities are set to $v_r = v_\theta = 0$. We assume an isothermal gas with sound speed $c_s = \sqrt{P/\rho} = 0.1 v_{\rm kep}(r=R_\star)$.

The magnetic field is set via a vector potential,
\begin{equation}
  \begin{aligned}
  A_\varphi = & v_{\rm A,0}\frac{R_\star^3}{2 {\rm \pi}}  \sqrt{\rho_0} \frac{\sin(\theta)}{r^2} \sin\left(2 {\tt \pi} \frac{r}{R_\star}\right) \\
  &\times \exp\left\{-\frac{1}{4} \left[\frac{\Omega_\star R_\star}{c_{\rm s}}\cos(\theta)\right]^2\right\},
  \end{aligned}
\end{equation}
where $v_{\rm A,0} = 0.002 v_{\rm kep}(r=R_\star)$ is the initial Alfv{\'e}n speed in the unperturbed star at the midplane. The magnetic field resulting from the curl of this vector potential is self-contained within the star (i.e., the inner boundary) and traces out concentric loops elongated in the radial direction and focused near the midplane where the density is peaked. The Alfv{\'e}n speed at $r=R_\star$ is largest at the midplane and then decreases as $\sin(\theta)$ with increasing $|\theta-{\rm \pi}/2|$.  Outside the star, the gas is set to the density floor, $\rho_{\rm flr} = 10^{-8} \rho_0 (r/R_\star)^{-3.5}$, with zero velocity and magnetic field. The outer radial boundary of the simulations is set to ``outflow'' boundary conditions, where the primitive variables are copied into the nearest ghost zones unless $v_r<0$, in which case the boundary is set to $v_r=0$.  The $\theta$ boundaries use ``polar'' boundary conditions \citep{Athena++} and the $\varphi$ boundaries are periodic.

When the density in a cell is $\le 50 \rho_{\rm flr}$, we impose a velocity ceiling of $|\mathbf{v}| = \sqrt{v_r^2 + v_\theta^2 + v_\varphi^2} \le \sqrt{2} v_{\rm kep}(r=R_\star)$ by reducing the magnitude of each component while maintaining the ratios between them (i.e., the overall direction).
For the MHD run, we impose additional floors on the density that depend on the local magnetic field strength. We do this by limiting the Alfv{\'e}n speed, 
\begin{equation}
  v_{\rm A}  \le  \min\left\{v_{\rm A,1}, v_{\rm A,2},v_{\rm A,3}\right\},  
  \end{equation}
  where 
  \begin{equation}
    \begin{aligned}
      v_{\rm A,1} &= 2.45 (2^{5-n}) \frac{r}{R_\star} \sin(\theta) v_{\rm kep}(r=R_\star)  \\  
      v_{\rm A,2} &= 1.52 (2^{5-n}) \frac{r}{R_\star} v_{\rm kep}(r=R_\star),  \\  
       v_{\rm A,3}  &= 5\sqrt{2} v_{\rm kep}(r=R_\star) ,
    \end{aligned}
  \end{equation}
   and $n$ is the local level of refinement.
These extra floors/limits in low density regions keep the Courant time step restriction at reasonable values without impacting the regions where there is a significant amount of matter.  We track the total amount of mass added into the simulations via floors and discuss them in more detail in Appendix \ref{App:floor}.

The domain extends from $ R_\star \le$ $r$ $\le 500 R_\star$, $0$ $\le$ $\theta$ $\le$ ${\rm \pi}$, and $0$ $\le$ $\varphi$ $\le$ $2 {\rm \pi}$.  
We adopt a base resolution of $64 \times 32 \times 12$ in $r,\theta$, and $\varphi$, with 5 levels of static mesh refinement (SMR) for MHD and 4 levels of refinement for hydrodynamics.   These extra levels are concentrated within 1--2 scale heights of the midplane, as described detail in Appendix \ref{App:grid}.  
Since we are focused on the dynamics of the midplane and expect the polar regions to be mostly evacuated, having lower resolution in those regions is not essential.
The radial grid spacing is logarithmic, while the angular spacing is uniform in both the poloidal and azimuthal directions. 

We run one MHD simulation for $1.2 \times 10^4 t_{\star}$ and one hydrodynamics simulation for $1.9 \times 10^4$ $t_\star$.

\section{Analytic Considerations}
\label{sec:analytic}
The scale height of an isothermal, Keplerian disc is roughly $H \approx r  c_{\rm s} / v_{\rm kep}  = 0.1 R_\star (r/R_\star)^{3/2}$.  We thus resolve the scale height by $\approx$ 16 cells in $\theta$ at $r = R_\star$ and $\approx$ 92 cells in $\theta$ at $r = 2 R_\star$.  The most unstable MRI wavelength is $\lambda_{\rm MRI} \approx 2 {\rm \pi}/\Omega v_{\rm A} $, where $\Omega = v_\varphi/[r \sin(\theta)]$. For Keplerian rotation and using the initial Alfv{\'e}n speed, this gives 
\begin{equation}
  \lambda_{\rm MRI }\approx 0.013 R_\star \left(\frac{r}{R_\star}\right)^{3/2},
  \end{equation}
  which is small enough to fit in the disc but just large enough to be resolved by our grid.  For example, at $r = 2 R_\star$, for this Alfv{\'e}n speed we would have $\approx$ 11 cells per wavelength in the polar and radial directions. In terms of pressure, the initial magnetic field strength is dynamically unimportant, as the ratio between thermal and magnetic pressure is $\beta_0 = 2 c_{\rm s}^2/v_{{\rm A},0}^2 =5000$.

Finally, we can estimate the expected disc radius (i.e., the critical point radius) using the approximation of \citet{Krticka2011}, 
\begin{equation}
  \label{eq:r_disc}
  r_{\rm disc} \approx \frac{3W}{10} \left(\frac{GM_\star}{ c_{\rm s}^2}\right)  \approx  27 R_\star 
\end{equation} 
for our simulation parameters.  This is safely well within the outer boundary of our domain.  In the $\alpha$-disc approximation, the viscous time is 
\begin{equation}
  \label{eq:t_visc}
t_{\rm visc} \approx \frac{r^2}{\alpha c_{\rm s} H} =\frac{100}{\alpha} \sqrt{\frac{R_\star^3}{GM_\star}}  \sqrt{ \frac{r}{R_\star}},
\end{equation}
which at $r = r_{\rm disc}$ and, gives $t_{\rm visc}(r=r_{\rm disc}) \approx 5200 (0.1/\alpha) t_\star. $ So we would expect our run time of $\approx 1.2 \times 10^4 t_\star$ for MHD to be sufficient for the entire disc to reach a steady state if $\alpha$ is $\gtrsim 0.04$.

\subsection{Typical Be Star Parameters}
\label{sec:units}
While the simulations we perform are scale-free, it is useful to convert our various parameters into physical units for a typical Be star.  For this purpose we use $\mu$ Centauri, which is a known pulsator \citep{Rivinius1998,Rivinius2001,Balona2001}.  

Using a combination of stellar atmosphere and stellar evolution modeling of observational data, \citet{Zorec2005} estimated the masses, surface gravity, and ages of nearly 100 galactic Be stars, including $\mu$ Centauri.   Their estimates were $M_\star \approx 9.3 M_\odot$, $R_\star \approx 5.3 R_\odot$, though the value for $R_\star$ depends on the assumed rotation rate of the star.  $\rho_0$ is more uncertain observationally, but fitting disc models to observed infrared excesses generally results in values $10^{-12}$--$10^{-10}$ g/cm$^3$ in Be stars (e.g., \citealt{Waters1987,Dougherty1994}). We thus adopt a fiducial value of $\rho_0 = 10^{-11}$ g/cm$^3$.   With these values we can convert parameters and simulation data to physical units, as we summarize in Table \ref{tab:parameters}. 

We use an artificially high temperature in order to be able to fully resolve the disc and the MRI.  Specifically, the observed effective temperature of $\mu$ Centauri is $T_{\rm eff} \approx 2.2 \times 10^4 K$, which is typical of Be stars \citep{Zorec2005}.  If the disc temperature is related to the effective temperature via $\sim$ $T_{\rm eff}/2$, then the $T = 41 \times 10^4 $ K used in our simulations is $\approx$ 37 times larger than appropriate.  We discuss the implications of this in \S \ref{sec:lim}.

\begin{table}
  \begin{center}
    \caption{Various parameters used in our simulations. We use the observed properties of $\mu$ Centauri to convert to physical units (see \S \ref{sec:units}). For the definition of each quantity see \S \ref{sec:methods}.}
    \label{tab:parameters}
    \def\arraystretch{1.75}
    \begin{tabular}{|c|c|} 
            \multicolumn{2}{c}{Stellar Parameters}\\
            \hline
                  $M_\star$ & $9.3 M_\odot$ \\
      \hline
       $R_\star$ & $5.3 R_\odot$ \\
       \hline
       $\rho_0$ & $10^{-11}$ g/cm$^3$ \\
       \hline
        $T$ & $41 \times 10^4$ K  \\
        \hline
        $c_{\rm s}$ & 58 km/s \\
        \hline
        $W$ & 0.95 \\
        \hline
      $\beta_0$ & 5000 \\
       \hline
       \multicolumn{2}{c}{Useful Quantities}\\
\hline
      $t_\star \equiv R_\star^3/(GM_\star) $ & 6.4 ks $\approx$ 0.07 d\\
      \hline
      $\rho_0 R_\star^3/t_\star $ & $1.2 \times 10^{-6}$ $M_\odot$/yr\\
      \hline
       $\rho_0 R_\star^3$ & $2.5 \times 10^{-10} M_\odot$\\
      \hline
      \multicolumn{2}{c}{Pulsation Parameters}\\
      \hline
      $v_{\varphi,{\rm pert}}/v_{\rm kep}(r=R_\star)$& 0.055 \\
      \hline
      $P$ & 26 ks $\approx$ 0.3 d\\
      \hline
      $m$ & 2 \\
      \hline
      $l$ & 2 \\
      \hline
      $ \varphi_0$ & 0\\
      \hline
      $\rho_{\rm max}/\rho_0$ & 1.1 \\
            \hline
    \end{tabular}
  \end{center}
\end{table}

\section{Results}
\label{sec:results}
\subsection{Comparing The Hydrodynamics and MHD Simulations}

\begin{figure*}
\includegraphics[width=0.49\textwidth]{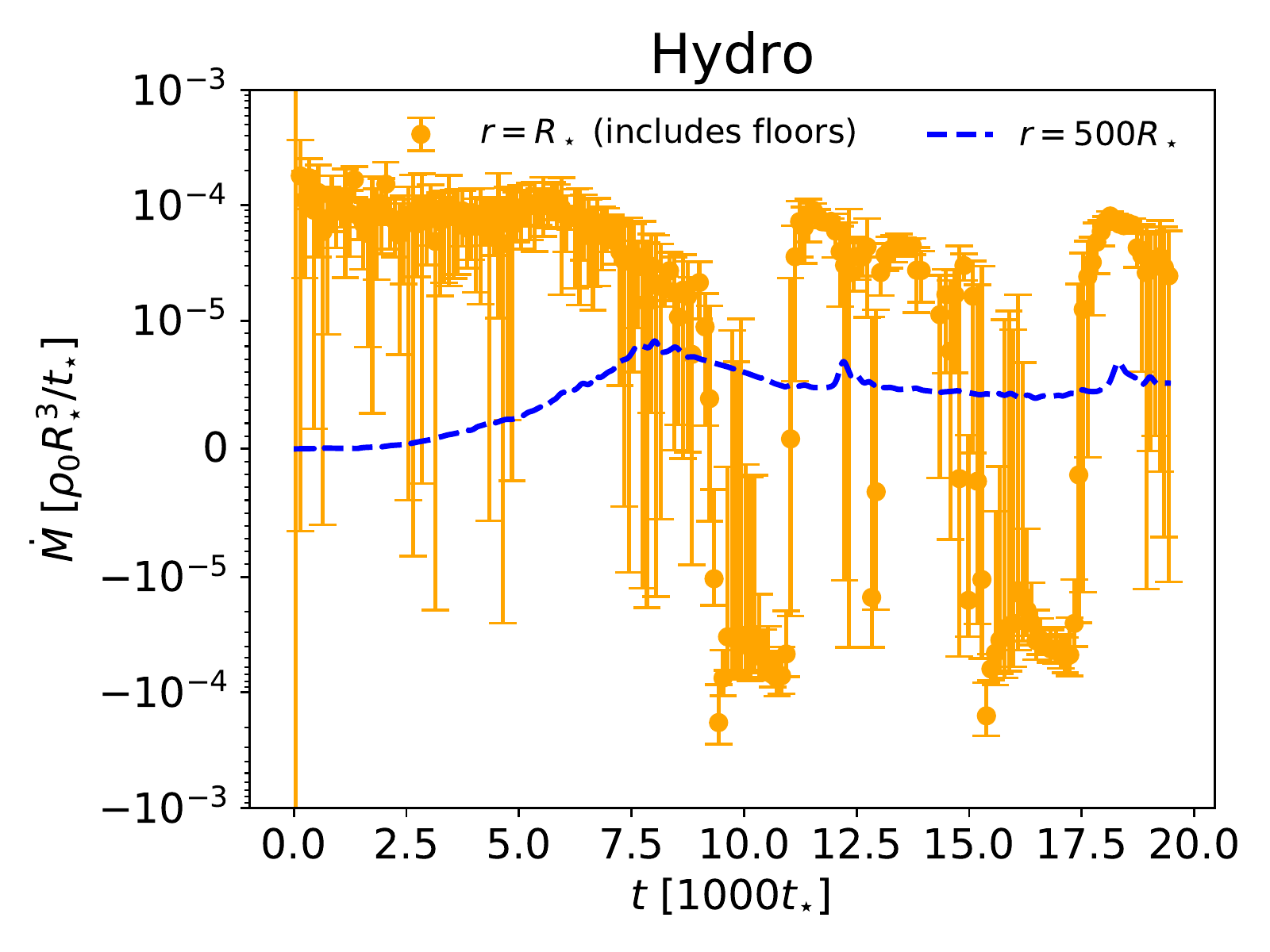}
\includegraphics[width=0.49\textwidth]{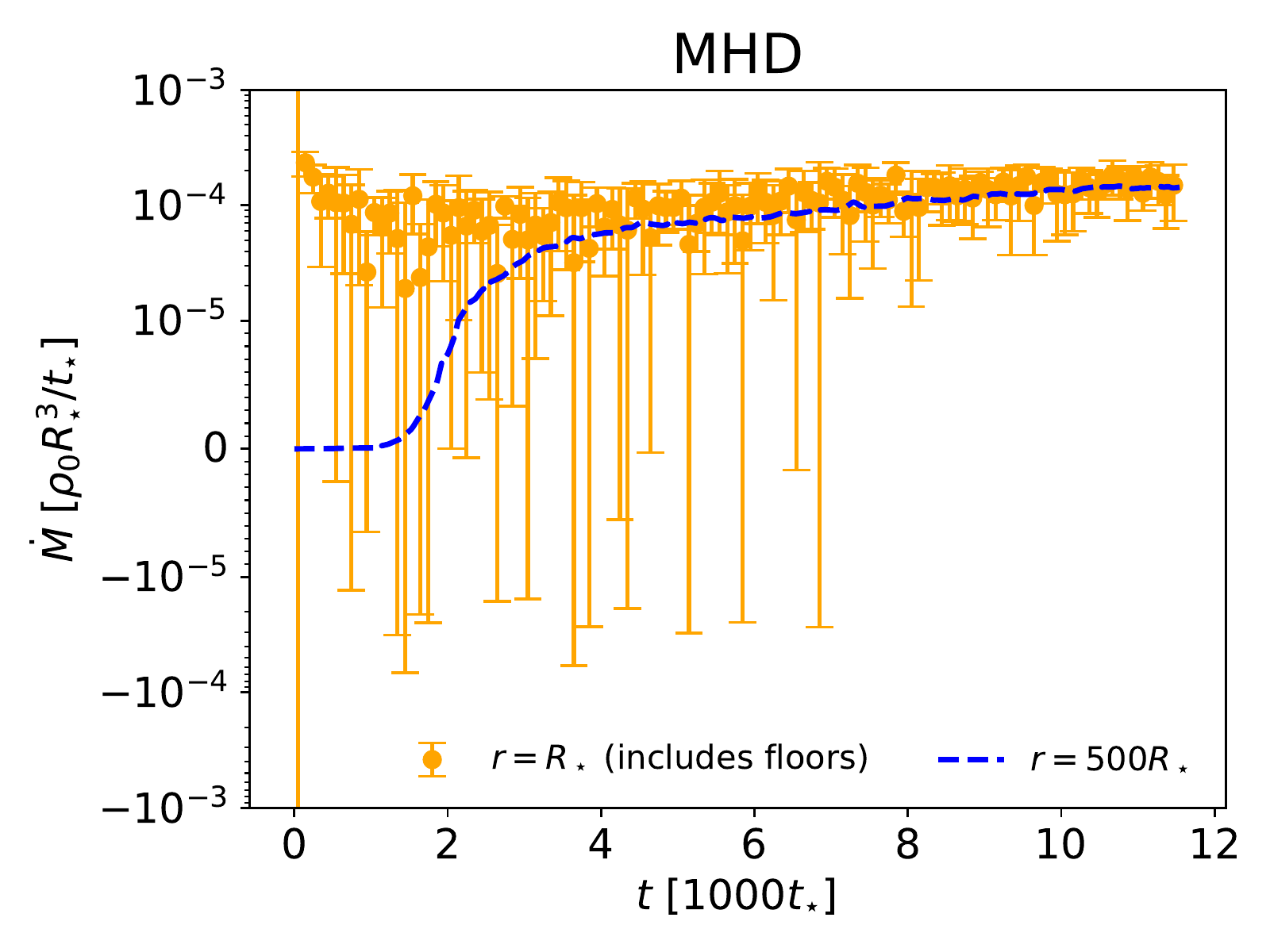}
\includegraphics[width=0.49\textwidth]{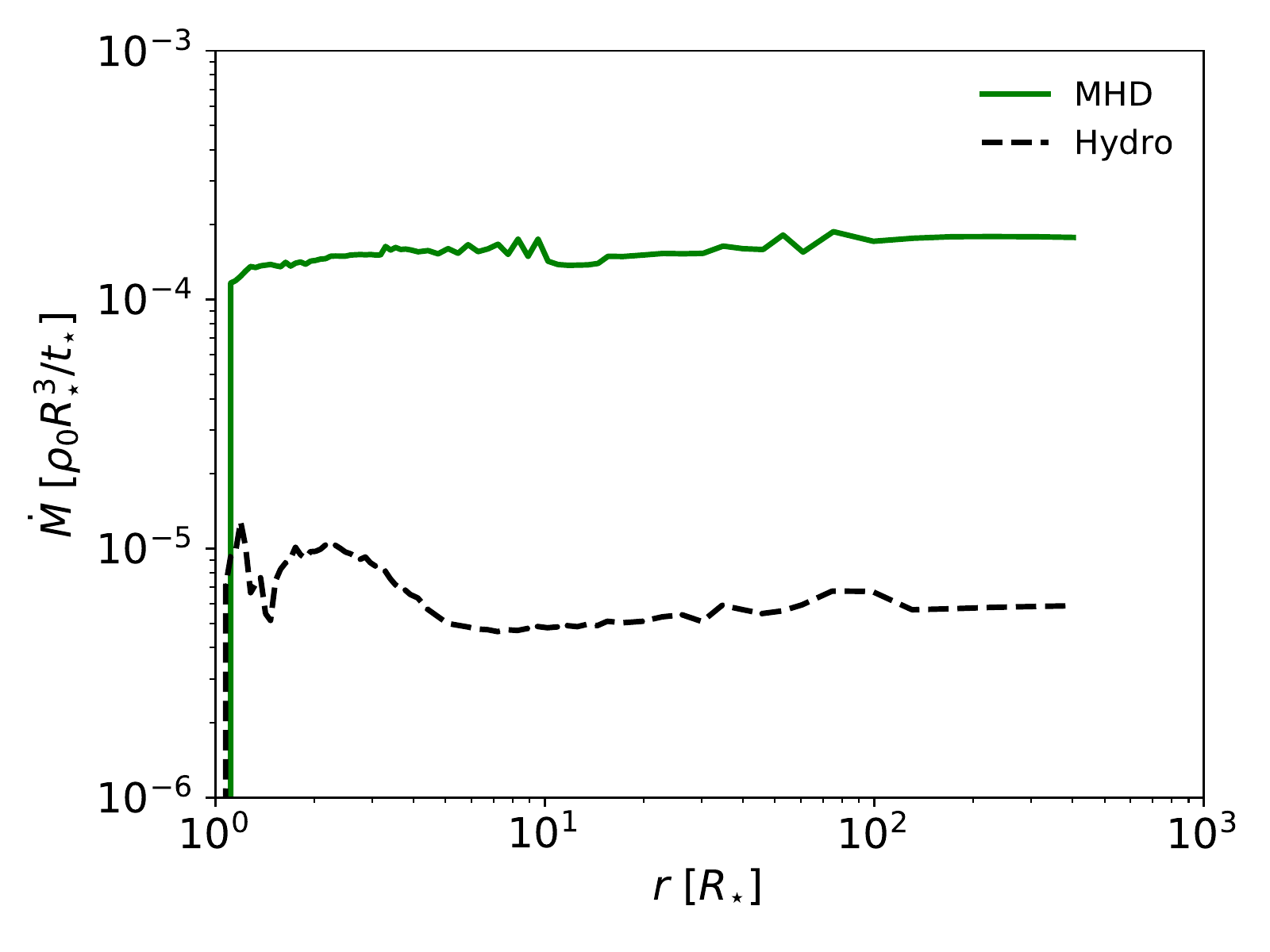}
\caption{  Comparison between decretion rates in the MHD and hydrodynamics simulations. Top: $\dot M$ measured through the inner boundary at $r= R_\star$ (orange circles) and the outer boundary (dashed blue) vs. time.  The left panel is the hydrodynamics simulations and the right panel is the MHD simulation.  Due to the rapid time variability, each data point for the $r=R_\star$ $\dot M$ has been averaged over 1000 $t_\star$, with the uncertainty representing one standard deviation.  Furthermore, $\dot M$ through the inner boundary includes the rate that mass is being added through the floors, which is negligible in the hydrodynamics simulation but significant in the MHD simulation.  This is because the floors generally add mass to the low density, highly magnetized polar regions just outside the inner boundary which then simply fall inwards through that boundary, as we show in Appendix \ref{App:floor}.   Bottom: $\dot M$ averaged over time vs. radius for the MHD (solid) and hydrodynamics (dashed) simulations. The graininess in these curves is due to the difficulty of calculating accurate angular integrals in post-processing for our complicated SMR grid (Figure \ref{fig:grid}).  Both simulations reach a form of steady state. In the hydrodynamics simulation this is because gas from the domain falls back through the inner boundary in cycles that results in the flux through the inner boundary oscillating in time from negative to positive, while in the MHD simulation this is because the flux through the outer boundary becomes large enough to balance the flux through the inner boundary.     } 
\label{fig:mdot}
\end{figure*}

The top panels in Figure \ref{fig:mdot} show the flux of mass through the inner and outer boundaries as a function of time for both the MHD and hydrodynamics simulations. Initially, both simulations have approximately the same $\dot M$ through the inner boundary, $\dot M_{\rm in} \approx 10^{-4} \rho_0 R_\star^3/t_\star$.  The precise value of this rate is a complicated function of the pulsation parameters, namely, $\rho_{\rm max}$, $v_{\varphi,{\rm pert}}$, $\varphi_0$, $m$, and $P$, as we have concluded after running several other low resolution and 2D test runs.  After $\sim$ 6000 $t_\star$, however, the mass flux through the inner boundary the hydrodynamics simulation decreases and even becomes negative.  This persists for a brief ($\lesssim 1000 t_\star$) interval during which matter is actually flowing out of the domain through the inner boundary.  Afterwards, $\dot M_{\rm in}$ rapidly returns to the original, positive value but then decreases and becomes negative once again and the cycle repeats.  As we shall see, this occurs because there is no efficient source of angular momentum transport; gas that is ejected from the ``star'' due to the imposed pulsations cannot escape to larger radii and a fraction of it falls back onto the surface. The mass flux through the outer boundary increases until it saturates at $\dot M_{\rm out} \approx 0.5 \times 10^{-5} \rho_0 R_\star^3/t_\star \ll |\dot M_{\rm in}|$. Due to the temporally varying sign of $\dot M_{\rm in}$, however, this results in an approximately steady state in which the total mass in the domain oscillates around a constant value.  In the MHD simulation, the picture is much different. Mass is steadily fed into the domain through the inner boundary at approximately the same rate at all times\footnote{This includes mass input from the floors, $\dot M_{\rm floor}$, which is significant.  In Appendix \ref{App:floor} we argue that $\dot M_{\rm floor}+\dot M_{\rm in}$ (the quantity shown in Figure \ref{fig fig:mdot}) is the appropriate quantity to plot given that the floors predominantly add mass in the polar regions at small radii which quickly fall through the inner boundary, decreasing the value of the naively calculated $\dot M_{\rm in}$.}, while the flux through the outer boundary increases until it is roughly equal to $\dot M_{\rm in}$, resulting in an approximately steady-state.  This can be seen in the bottom panel of Figure \ref{fig:mdot}, which plots the time-averaged decretion rate as a function of radius for the two simulations, where the MHD $\dot M$ is $\approx$ constant with radius.  The hydrodynamics $\dot M$ is $\approx$ constant for $r \gtrsim 5 R_\star$, but has a peak at about twice that value for smaller radii.  This is a result of averaging over an interval that contains both positive and negative $\dot M_{\rm in}$ but only includes a few cycles of oscillation.  Over a long enough timescale we expect that this peak in the time-averaged $\dot M$ vs. radius would not be present, showing $\dot M \approx$ ${\textrm const.}$ at a value equal to the saturated $\dot M_{\rm out}$ seen in the top left panel of Figure \ref{fig:mdot}.  The location of the peak is indicative of where the majority of the mass in the simulation is located.

\begin{figure*}
\includegraphics[width=0.95\textwidth]{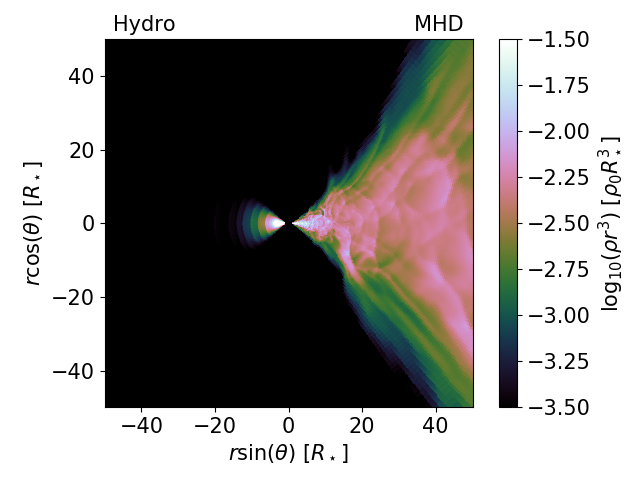}
\caption{Time slices of density weighted by $r^3$ at $t =  10^4 t_\star$ for the hydrodynamics (left) and MHD (right) simulations.  Because of the enhanced angular momentum transport provided by magnetic fields, the MHD simulation shows a more extended distribution of mass compared to the hydrodynamics simulation, where mass is concentrated at $r$ $\sim$ $5 R_\star$.  
} 
\label{fig:mhd_v_hydro_contour}
\end{figure*}

This is clear in Figure \ref{fig:mhd_v_hydro_contour}, which compares density contour time-slices at $t=10^4 t_\star$ and $\varphi=0$ in the two simulations, where the density has been scaled by a factor of $r^3$ in order to see detailed structure at all radii.  The MHD simulations shows a turbulent disc of gas that expands into a broader spherical distribution at $\sim$ 28--30 $R_\star$.  Mass is more or less evenly distributed across all radii.  The hydrodynamics simulation also shows a disc with the same scale height, but the flow is laminar and mass is concentrated at smaller radii, $\sim$ $5 R_\star$ (the same region with the peak in $\dot M$ shown in the bottom panel of Figure \ref{fig:mdot}).  

\begin{figure}
\includegraphics[width=0.45\textwidth]{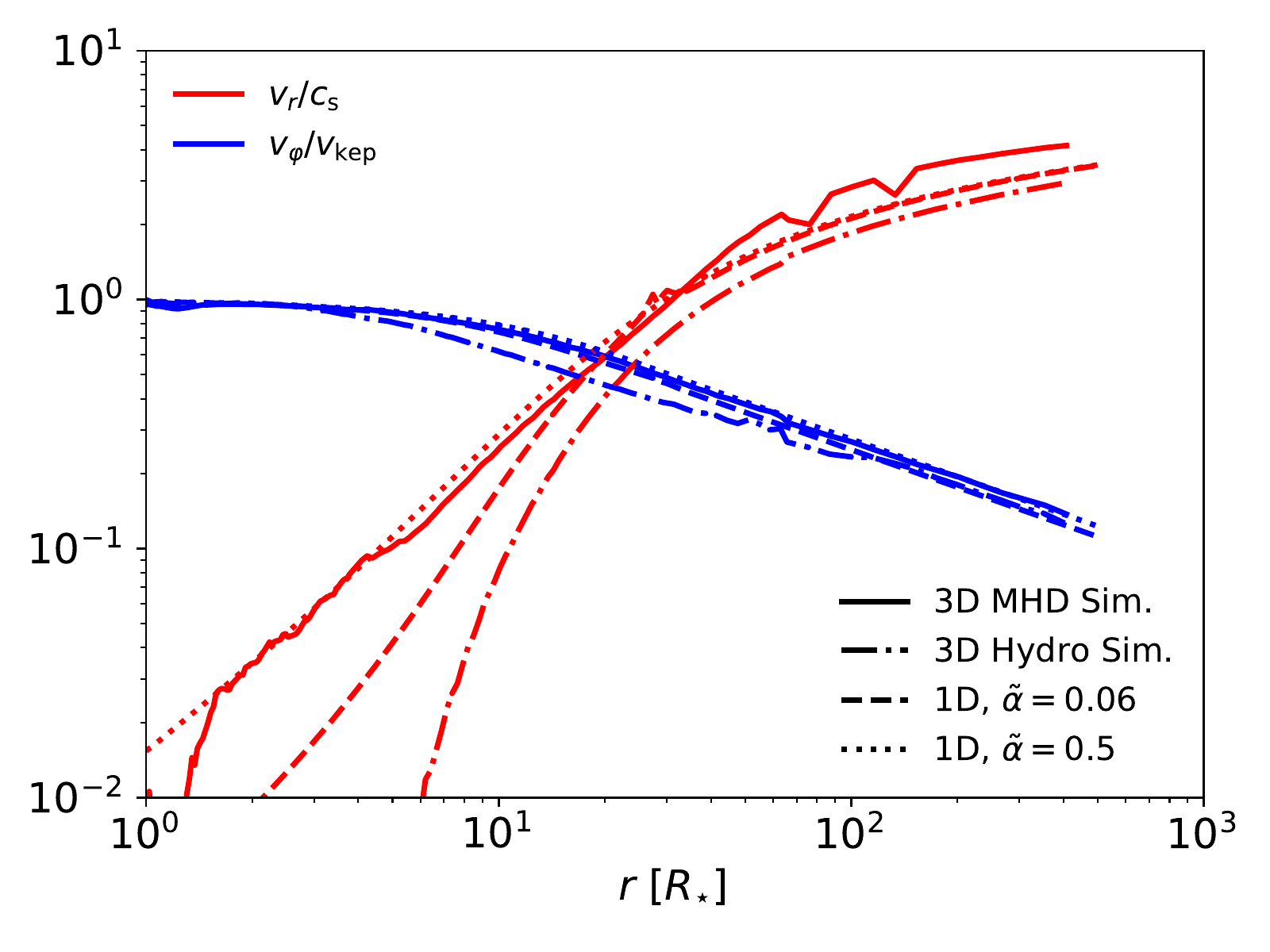}
\includegraphics[width=0.45\textwidth]{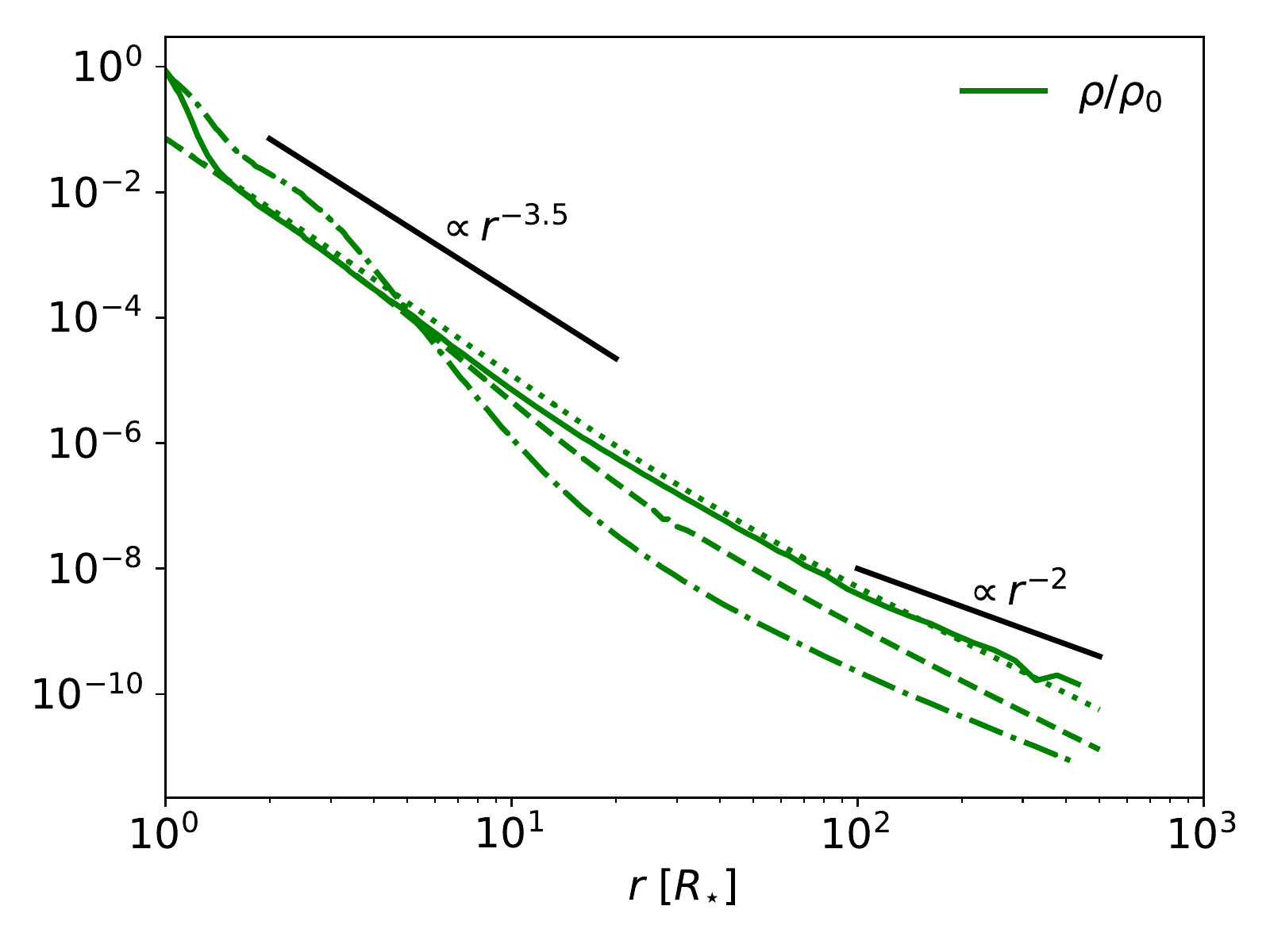}
\caption{ Quantities in our 3D simulations compared to the 1D model of \citeauthor{Okazaki2001} (\citeyear{Okazaki2001}, modified slightly as described in the text).  Solid lines are for the 3D MHD simulation, dot-dashed lines are for the hydrodynamics simulation, dashed line are for the 1D model with $\tilde \alpha =0.2$, and dotted lines are for the 1D model with $\tilde \alpha =1.0$.   Top: Radial velocity divided by the sound speed (blue), $\langle v_r/c_s \rangle_\rho$, and azimuthal velocity divided by the Keplerian speed (red), $\langle v_\varphi/v_{\rm kep}\rangle_\rho$. Bottom: Mass density, $\langle \rho/\rho_0 \rangle_{|r\cos(\theta)|<H}$.  Note that $\rho$ in the 1D model has been scaled to align with the 3D simulation since the density scale is arbitrary.  The 1D model matches the angle and time-averaged profiles of the MHD simulation very well for $\tilde \alpha = 1$, while the hydrodynamics simulation is not well represented by either 1D model.  In the MHD simulation and the 1D models, radial velocity increases as $\sim$ $r^{-1}$ until the sonic point at $\approx$ 28--30 $R_\star$, at which point it increases more gradually as it approaches an asymptotic limit. Correspondingly, in these same models the density decreases as $r^{-3.5}$ below the sonic point and approaches $r^{-2}$ at larger radii.  Azimuthal velocity is $\approx$ Keplerian until $\approx$ 10 $R_\star$, where it starts transitions to a constant angular momentum profile of $v_\varphi \propto r^{-1}$ at $r\gtrsim 30 R_\star$.  The profiles of the hydrodynamics simulation differ from the rest in that they show an accumulation of density at $r \approx 2$--4$R_\star$ and a much lower radial velocity below the sonic point.  The sonic point is also much further out, $\approx$ 40 $R_\star$.  }
\label{fig:3D_1D_comp}
\end{figure}

Due to our assumption of an isothermal flow, the scale height ratio $H/r$ increases with radius so that the disc expands in the $z$ direction as it decretes outwards.  It starts out relatively thin, with $H/r$ $\approx$ 0.1, but by $\approx$ 20 $R_\star$ it reaches $H/r$ $\sim$ 1 as seen in Figure \ref{fig:mhd_v_hydro_contour}. At this point it no longer resembles a disc but a spherical wind.  The transition between disc and wind can be measured more precisely as the location of the sonic point, where the flow transitions from subsonic to supersonic.  In MHD this occurs at $\approx$ 28--30 $R_\star$ (in excellent agreement with the estimate of $\approx$ 27 $R_\star$ in Equation \ref{eq:r_disc}), while in hydrodynamics it occurs at a larger radius $\approx 40 R_\star$, as shown in Figure \ref{fig:3D_1D_comp}, which plots the mass-weighted and angle-averaged radial velocity relative to the sound speed as a function of radius.   In the subsonic region, $v_r$ rises as $\tilde \propto$ $r$ for MHD and much steeper in hydrodynamics (where the magnitude of $v_{r}$ is also much smaller), while in the supersonic regime it increases more slowly with radius as it approaches an asymptotic value in both simulations.  A transition is also seen in $v_\varphi$ at roughly the same radius in MHD; in the subsonic region $v_\varphi/v_{\rm kep}$ $\approx$ 1, while in the supersonic regime $v_\varphi/v_{\rm kep}$ $\tilde \propto$ $r^{-1/2}$.  In other words, the gas transitions from being roughly Keplerian to roughly angular momentum conserving at the sonic point.   This is consistent with a transition from a thin disc to a thick wind.  Hereafter we refer to gas in the subsonic region of the MHD simulation ($r\lesssim 30 R_\star$) as the `disc' and gas in the supersonic regime ($r \gtrsim 30 R_\star$) as the `wind.'  In hydrodynamics, the transition to an angular momentum conserving flow occurs at a smaller radius, $r \approx 4 R_\star$. 

\begin{figure*}
  \begin{centering}
\includegraphics[width=0.95\textwidth]{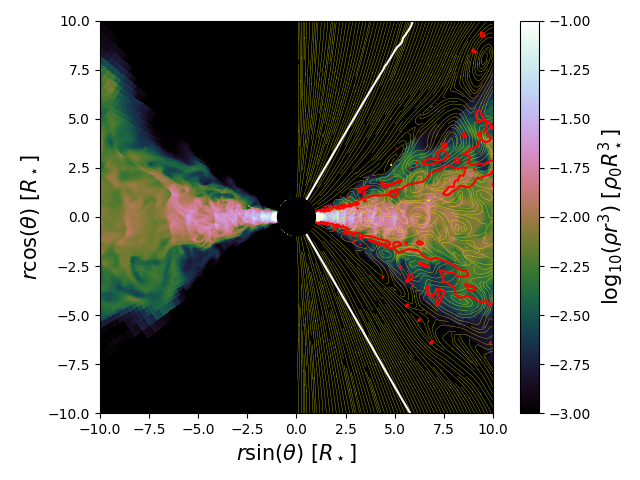}
\end{centering}
\caption{ 2D poloidal slice of $\rho r^3$ at $t = 16000 t_\star$ in the MHD simulation. Magnetic field lines are overplotted in the right panel in yellow.  Red lines mark the contour of $\langle B_r\rangle \langle B_\varphi\rangle/\langle B_r B_\varphi \rangle = 0.5$, while white lines mark the contour of $\arccos(|\mathbf{B} \cdot \mathbf{\hat s}|/|\mathbf{B_{\rm p}}|) =60^\circ$, where $\mathbf{\hat s}$ is the unit vector in the direction of the cylindrical radius and $\mathbf{B_{\rm p}}$ is the poloidal magnetic field.  
The decretion disc is clearly turbulent, which is caused by the MRI.  The field lines transition from being turbulent in the midplane to strongly coherent in the poles.  The region between the white and red contours is where coherent magnetic torques efficiently operate \citep{BP1982}.
  } 
\label{fig:mhd_contour}
\end{figure*}

\begin{figure}
\includegraphics[width=0.45\textwidth]{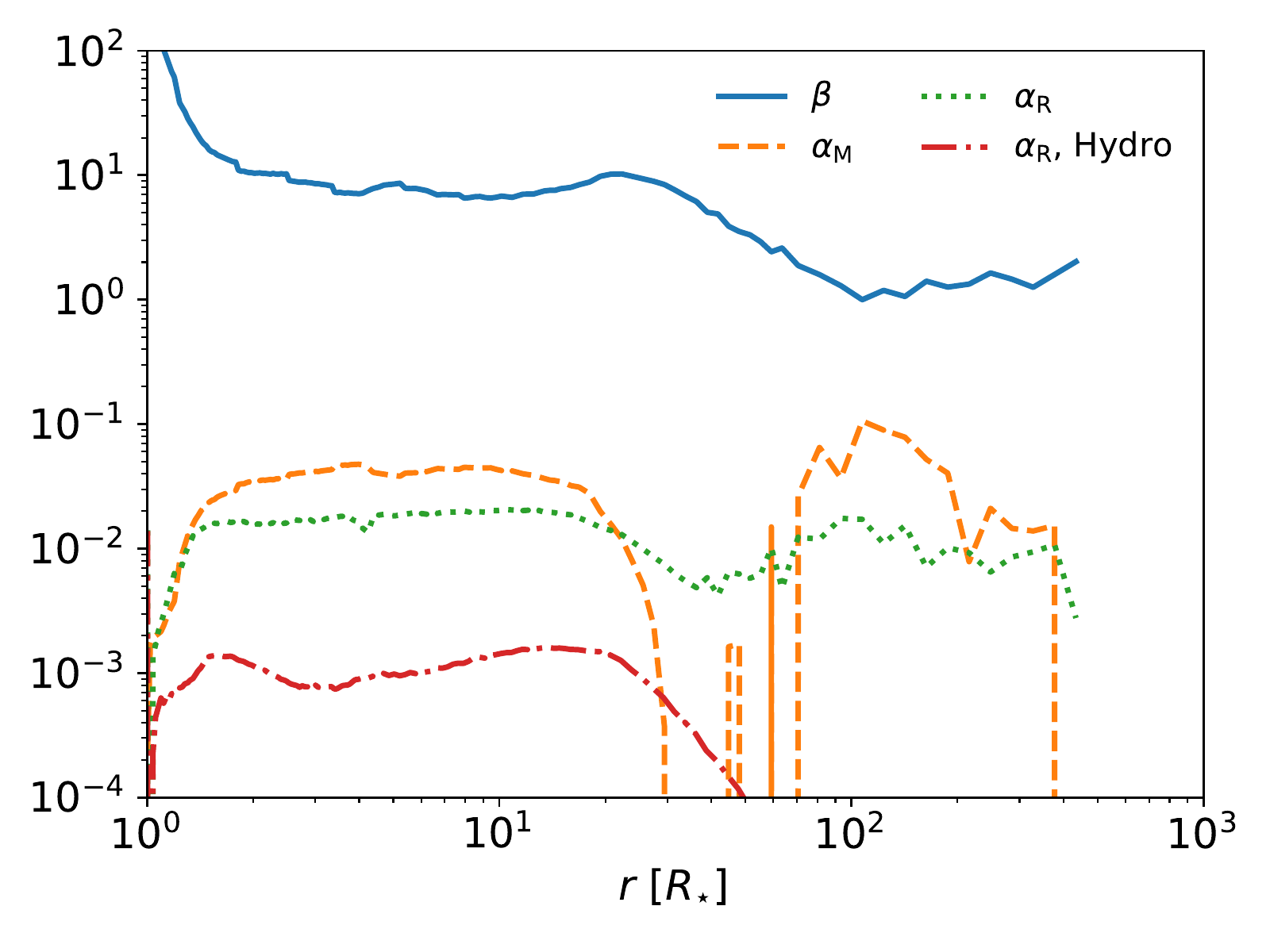}
\caption{Time and angle-averaged $\beta$, $\langle \beta^{-1} \rangle_\rho^{-1}$, as well as Maxwell and Reynolds stresses, $\alpha_{\rm M}$ and $\alpha_{\rm R}$ (Equations \ref{eq:alpha_m} and \ref{eq:alpha_r}), plotted vs. radius.  These quantities are evaluated only for $|r \cos(\theta) | <H|$, where $H$ is the density scale height.  In the MHD simulation, the MRI amplifies the magnetic field from $\beta_0 =5000$ to $\beta \lesssim 10$ by $r =  2 R_\star$, leading to an $\alpha_{\rm M}$ of $\approx$ 0.04--0.05.  The Reynolds stress is a factor of $\approx$ 2--3 smaller than the Maxwell stress below the sonic point radius.  In the hydrodynamics simulation, the Reynolds stress is significantly smaller, $\approx$ 10$^{-3}$ below the sonic point radius and negligible at larger radii.
} 
\label{fig:alpha_beta}
\end{figure}

In addition to the build up of mass at small radii in hydrodynamics, the other obvious difference between the two simulations in Figure \ref{fig:mhd_v_hydro_contour} is the presence of turbulence in MHD. Figure \ref{fig:mhd_contour} shows a smaller scale ($20 R_\star \times 20 R_\star$) slice of density (again scaled by $r^3$) overplotted with magnetic field lines in the MHD simulation to highlight the turbulent structure of the disc.  This turbulence is caused by the MRI, which amplifies the initial magnetic field from $\beta_0 =5000$ to $\beta \approx 10$ just outside the star, as shown in the time and angle-averaged $\langle \beta^{-1} \rangle_\rho^{-1}$ vs. radius plotted in Figure \ref{fig:alpha_beta} (note that the averages are restricted to within one scale height off the midplane).  Within the disc, $\beta$ is, on average, $\lesssim 10$ and only a weakly increasing function of radius. 
In the wind, $\beta$ increases drops to $\approx$ 2 by $100 R_\star$.  The magnetic field in the disc is predominantly in the $\varphi$ direction, with $\sqrt{\langle B_r^2 + B_\theta^2\rangle_\rho/\langle B_\varphi^2 \rangle_\rho} \approx$ 0.4--0.5 below the critical point, corresponding to a root-mean-squared angle of $\approx$ 20$^\circ$ with respect to the $\varphi$ direction, consistent with local shearing box simulations of the MRI (e.g., \citealt{Guan2009}).  Above the critical point $\sqrt{\langle B_r^2 + B_\theta^2\rangle_\rho/\langle B_\varphi^2 \rangle_\rho}$ declines as $\tilde \propto$ $r^{-0.75}$ (that is, the field becomes more and more toroidal with increasing radius).

\begin{figure*}
  \begin{centering}
    \includegraphics[width=0.85\textwidth]{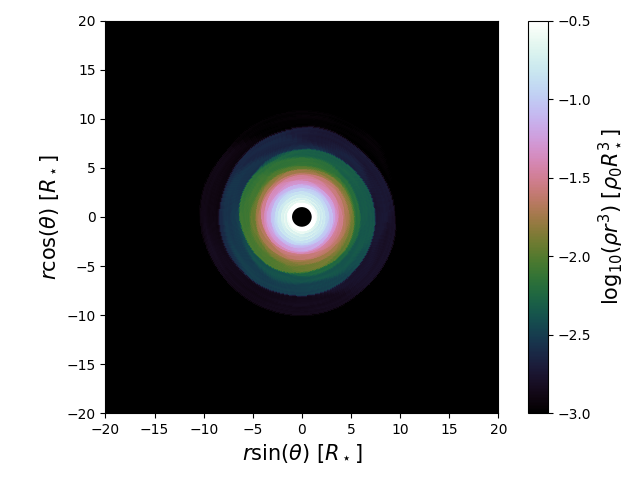}
\includegraphics[width=0.85\textwidth]{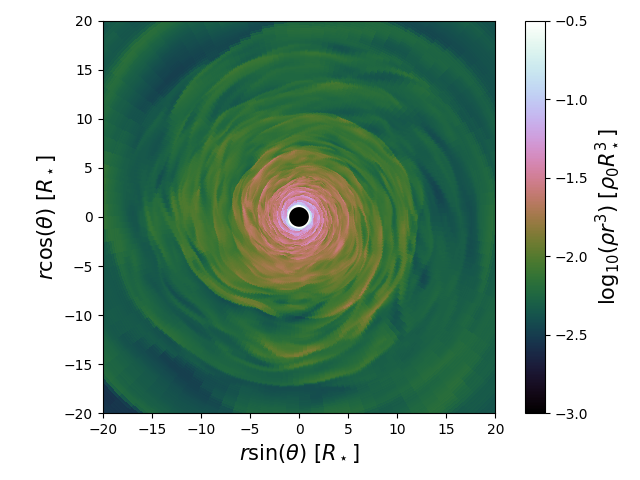}
\end{centering}
\caption{ 2D $\theta=\pi/2$ slice of $\rho r^3$ at $t = 10^4 t_\star$ in the hydrodynamics (top) and MHD (bottom) simulations. Spiral structures are seen in the MHD simulation that propagate outwards but they are incoherent in space and time while contributing only marginally to angular momentum transport ($\alpha_{\rm m}/\alpha_{\rm r} \sim$ 3--4 (Figure \ref{fig:alpha_beta}).  The hydrodynamics simulation is more axisymmetric, with the gas having circularized after being confined to small radii due to the lack of significant angular momentum transport.} 
\label{fig:midplane_contour}
\end{figure*}

The $r$--$\varphi$ stress caused by the magnetic field can be quantified using the dimensionless \citet{Shakura1973} parameter (e.g., \citealt{Jiang2019}):
\begin{equation}
  \label{eq:alpha_m}
  \alpha_{\rm M} = \frac{\langle -B_r B_\varphi \sin(\theta) \rangle }{\langle P + P_{\rm M} \rangle},
\end{equation}
where $P_{\rm M}$ is the magnetic pressure.  Similarly, the hydrodynamic Reynolds stress can be quantified using 
\begin{equation}
   \label{eq:alpha_r}
  \alpha_{\rm R} = \frac{\langle \rho v_r v_\varphi \sin(\theta)\rangle - \langle \rho v_r \rangle \langle v_\varphi \sin(\theta) \rangle }{\langle P + P_{\rm M} \rangle}.
\end{equation}
Both are plotted vs. radius in Figure \ref{fig:alpha_beta} calculated for $|r \cos(\theta)|<H$.  For MHD, within the disc region both $\alpha_{\rm M}$ and $\alpha_{\rm R}$ are approximately constant and equal to $\approx$ 0.04--0.05 and 0.01--0.02, respectively, while in the wind region their behavior is more complicated but less relevant because the flow is supersonic and angular momentum is no longer dynamically important.  This Maxwell stress is dominated by localized, turbulent correlations instead of global, mean-field torque; i.e., $|\langle B_r B_\varphi\rangle| \gg |\langle B_r\rangle \langle B_\varphi\rangle|$, at least for the $|r \cos(\theta)|<H$ region. For hydrodynamics, $\alpha_{\rm R}$ is much lower, $\approx$ $10^{-3}$ for $r\lesssim 20 R_\star$, after which it decreases with radius.  This lack of angular momentum transport is why mass cannot efficiently reach $r \gtrsim 5 R_\star$ in the hydrodynamics simulation (Figures \ref{fig:mhd_v_hydro_contour}) and why the decretion rate at large radii is a factor of $\sim$ 20 smaller than the MHD simulation (Figure \ref{fig:mdot}).  
To put these values of $\alpha_{\rm M}$ and $\alpha_{\rm R}$ in context, local shearing box simulations of MRI accretion discs find that the magnitude of the Reynolds stress is typically 1/4 the Maxwell stress
(\citealt{Pessah2006} and references therein).  In global simulations, however, the relative amount of Reynolds stress can be larger and even dominate the Maxwell stress if spiral density waves are excited (e.g., by a companion star \citealt{Ju2017} or by dynamically important radiation \citealt{Jiang2019}). Here and throughout we use `spiral density waves' to describe waves (or really, shocks) that have both a radial and azimuthal component to their velocity.     Such waves generally transport angular momentum outwards by effectively reducing the azimuthal velocity downstream of the shock fronts.   
In our simulations, $|\alpha_{\rm m}/\alpha_{\rm r}| \approx $ 3--4, more or less consistent with the shearing box simulations.  While we do see non-axisymmetric, spiral-like structures in the MHD disc (bottom panel of Figure \ref{fig:midplane_contour}, which plots a midplane slice of $\rho r^3$ at $t = 10000t_\star$), these structures are not coherent.  When attempting to isolate specific arms we find that they are not well fit by the linear dispersion relation that relates $r$ to $\varphi$  \citep{Binnney2008,Jiang2019}).   The hydrodynamics simulation is more or less axisymmetric, as seen in the top panel of Figure \ref{fig:midplane_contour}.

In an attempt to quantify the variability associated with spiral waves we also calculated the temporal Fourier transform of the midplane, $\varphi=0$ density at several different radii.  
Only the smallest radii ($r\lesssim 3 R_\star$) show show clear peaks in the spectrum, located at the harmonics of the pulsation frequency ($1/P$).  The spectrum at larger radii is dominated by turbulent variability that shifts to lower frequencies as radius increases.  This is expected based on the argument of \citet{Kato1983}, which states that in cold, nearly Keplerian discs, only very low frequency $m=1$ modes can be coherent across different radii (see also \citealt{Okazaki1991}, who first applied the argument to Be discs). This is because the frequency of the waves must obey $\omega - m \Omega(r) \approx \kappa(r)$, where $\omega =2{\rm \pi}/P$, $\Omega = v_\varphi/[r\sin(\theta)]$, and $\kappa$ is the epicyclic frequency, which for Keplerian rotation is equal to $\Omega$.  Thus, in order for the wave to be coherent (i.e., $\omega$ independent of $r$), $m$ must be equal to 1 and $\omega \ll \Omega$ (with the precise value determined by the small difference between $\kappa$ and $\Omega$ when the rotation is not exactly Keplerian).  Since the waves excited by the pulsating inner boundary have $m=2$ and $\omega$ comparable to $\Omega$, they cannot propagate coherently to larger radii.  Practically this means their effect on angular momentum transport is limited.

\subsection{Comparison to 1D Models}
 We now compare time and angle-averaged properties of our MHD simulation to 1D models typically used in the literature.  In particular, \citet{Okazaki2001} derived solutions for an $\alpha$-disc, isothermal decretion flow: 
\begin{equation}
  \label{eq:1D_model}
  \begin{aligned}
    \Sigma v_r r &= \textrm{const.} \\
    v_\varphi &=  \frac{l_s}{r}  + \frac{\alpha c_{\rm s}^2} {r} \left(\frac{r_{\rm s}}{c_{\rm s}} - \frac{r}{v_r}\right)\\
    v_r \frac{{\rm d}v_r}{{\rm d}r} &= \frac{v_\varphi^2}{r} - \frac{GM}{r^2}  - \frac{1}{\Sigma} \frac{ {\rm d} \left(c_{\rm s}^2 \Sigma \right)}{{\rm d}r}  + \frac{3}{2} \frac{c_{\rm s}^2}{r},
  \end{aligned}
\end{equation}
where $r_{\rm s}$ is the sonic radius, $l_{\rm s}$ is the specific angular momentum at $r_{\rm s}$, and $\Sigma$ is the surface density, which is related to the volume density by $\Sigma = \sqrt{2 {\rm \pi}} \rho r c_{\rm s}/v_{\rm kep}(r)  $.  We modify the model slightly from the original paper by making $\alpha$ a step function,
\begin{equation}
\alpha =    \left\{
 \begin{array}{ll}
     \tilde \alpha & r\leq r_{\rm s} \\
      0& r>r_{\rm s}.  \\
\end{array} 
\right. 
\end{equation}
Setting $\alpha$ to 0 for $r>r_{\rm s}$ better represents the loss of viscous coupling that occurs in our simulations once the flow becomes super-sonic. In Equation \eqref{eq:1D_model}, the solutions for $v_r$ and $v_\varphi$ are independent of the density scale, while $\Sigma$ scales uniformly with $\Sigma_\star = \Sigma (r = R_\star)$.   We solve these equations using the shooting method (see \citealt{Okazaki2001} or \citealt{Krticka2011} for more details) and plot the results for $\tilde \alpha=0.5$ and $\tilde \alpha =0.06$ compared to our 3D MHD and hydrodynamics simulations in Figure \ref{fig:3D_1D_comp}.  While the value of $\tilde \alpha = 0.06$ better represents the total $\alpha$ computed from our MHD simulations (Figure \ref{fig:alpha_beta}), the 1D model with $\tilde \alpha =0.5$ better represents the MHD radial profiles.  In fact, the agreement with the MHD simulation for $\tilde \alpha =0.06$ is excellent, with the the biggest difference being that the simulation has a $\sim$ 50\% higher radial velocity at large radii ($r \gtrsim 30 R_\star$).
The $\tilde \alpha =0.06$ model, on the other hand, underproduces the radial velocity by a factor of $\sim$ 2--4 for $r \lesssim 10 R_\star$.  
The fact that the $\tilde \alpha=0.5$ model better represents our MHD simulation hints that there may be interesting angular structure in the simulations that is not captured in the 1D model.  We confirm this in the next subsection.
The hydrodynamics simulation curves in Figure \ref{fig:3D_1D_comp} are not well fit by either 1D solution shown, but rather are closer to an $\tilde \alpha \ll 0.06$ solution with much smaller radial velocities than the $\tilde 
\alpha = 0.06$ and $\tilde \alpha =0.5$ curves.   This makes sense since $\alpha_{\rm R}$ is $\lesssim 10^{-3}$ for all radii in the hydrodynamics simulation (Figure \ref{fig:alpha_beta}).

  
  \subsection{Angular Structure of the MHD Simulation}
\begin{figure*}
\includegraphics[width=0.95\textwidth]{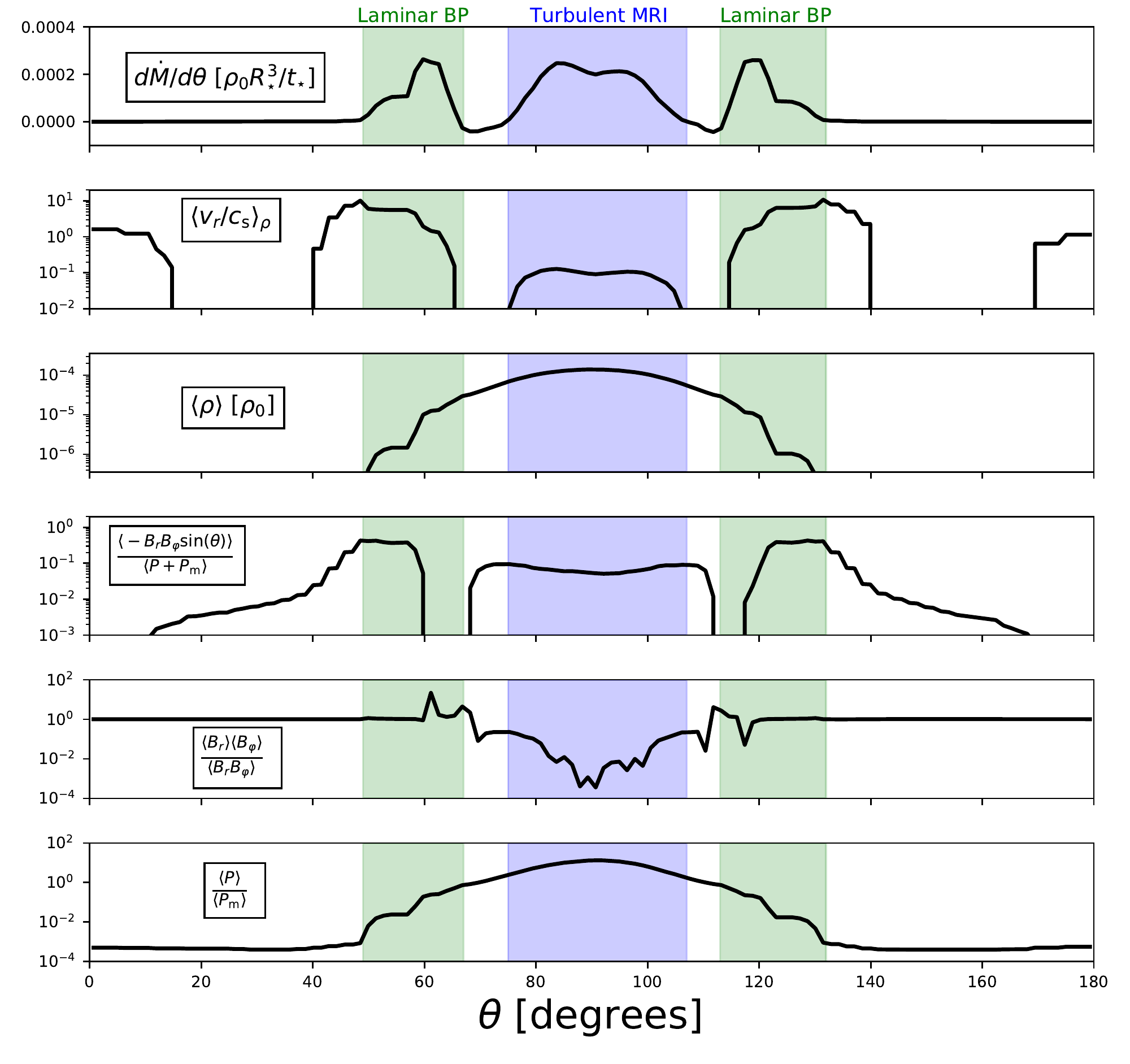}
\caption{Various time and $\varphi$-averaged quantities plotted vs. $\theta$ at $r=5 R_\star$, including the distribution of decretion rate with polar angle, $d \dot M/d \theta$ (top), radial Mach number, $\langle v_r/c_{\rm s}\rangle_\rho$ (2nd panel), mass density, $\langle \rho \rangle$ (3rd panel), Maxwell stress, $\langle - B_r B_\varphi \sin(\theta) \rangle/\langle P+P_{\rm m} \rangle$ (4th panel), a measure of magnetic field coherence, $\langle B_r\rangle \langle B_\varphi\rangle/\langle B_r B_\varphi\rangle$ (5th panel), and plasma $\beta$, $\langle P \rangle /\langle P_{\rm m}\rangle$ (bottom panel).  Decretion primarily happens in two distinct regions: the midplane of the disc (shaded blue) and the surface of the disc (shaded green).  In the high density, $\beta \lesssim 10$ midplane, the turbulent MRI stress mediated by disordered magnetic fields drives a highly subsonic outflow.  In the low density, $\beta \sim 10^{-1}$ surface regions, ordered magnetic fields drive supersonic outflow via the \citeauthor{BP1982} (\citeyear{BP1982}, BP) process. } 
\label{fig:theta}
\end{figure*}

Our analysis thus far has focussed primarily on the region of our simulations within one scale height of the disc midplane.   We now show that the MHD simulation has significant angular structure not captured by 1D models.  Figure \ref{fig:theta} shows six different time and azimuthally-averaged quantities vs. polar angle, including the decretion rate, the radial Mach number, the mass density, the Maxwell stress, a measure of how ordered the magnetic field is ($\langle B_r \rangle\langle B_\varphi\rangle /\langle B_r B_\varphi\rangle$), and plasma $\beta$.  Decretion is seen to be significant not just near the midplane, but also on the surface of the disc.  The midplane decretion is driven by MRI turbulent stresses, with highly disordered magnetic fields, high density, slow radial velocity, and Maxwell stresses $\alpha_{\rm M} \lesssim 0.05$.  Conversely, the surface decretion is driven by strong \citeauthor{BP1982} (BP, \citeyear{BP1982}) stresses ($\alpha_{\rm M} \gtrsim 0.5$) mediated by a global, coherent magnetic fields with $\langle B_r \rangle\langle B_\varphi\rangle /\langle B_r B_\varphi\rangle \approx 1$ and $\beta \sim 10^{-2}$--1.  As a result, this region displays supersonic velocities ($\langle v_r/c_{\rm s}\rangle \lesssim 10$), which results in significant decretion despite the density being $\lesssim 10$ times lower than the midplane.  

The \citet{BP1982} process is typically described using the analogy of a beads on a wire, where lower density plasma (the ``beads'') is centrifugally accelerated by magnetic field lines (the ``wires'').  Such centrifugal ejection from the disc is possible only if the magnetic field is coherent, strong (i.e., relatively low $\beta$), and if it makes less than an angle of $60^\circ$ with the surface of the disc. All three of these conditions are met in the surface outflow seen in our MHD simulation in Figure \ref{fig:theta}.   In this region, $\beta \lesssim 1$, $\langle B_r \rangle\langle B_\varphi\rangle /\langle B_r B_\varphi\rangle \approx 1$, and the surface of the disc has a coherent poloidal field close to parallel with the disc (in fact, making an angle of $<60^\circ$ even with the midplane), as seen in Figure \ref{fig:mhd_contour}, where contours of $\langle B_r \rangle\langle B_\varphi\rangle /\langle B_r B_\varphi\rangle=0.5$ and $\arccos(|\mathbf{B} \cdot \mathbf{\hat s}|/|\mathbf{B_{\rm p}}|) =60^\circ$ are shown.   All of this together gives strong evidence that the  \citet{BP1982} process is driving the surface decretion in our MHD simulation.

We should note that the transition between the turbulent MRI region and the laminar BP region occurs very close to a SMR refinement boundary (located at $\theta = 3 {\rm \pi}/8$, $5 {\rm \pi}/8$) where the resolution drops by a factor of 2. In fact, the laminar BP region tends to overlap with several refinement boundaries in the transition from higher to lower resolutions.  This is most likely correlative and not causative, as by design we constructed the grid to best resolve 1--2 scale heights from the midplane where the density is highest.  This means that the refinement boundaries are located where the density is steeply declining and the flow is becoming more magnetically dominated, precisely the location where any BP-type process would be expected to drive winds.
  
\section{Limitations of Our Simulations}
\label{sec:lim}
Our model of feeding the disc from the stellar surface relies on the pulsation driven orbital mass ejection paradigm for decretion discs.  While this model is generally considered a promising explanation for the Be disc phenomenon, it is by no means the only feasible mechanism for mass ejection, as discussed in the introduction.  Furthermore, we have assumed that pulsations persist indefinitely, resulting in a steady state disc solution (at least in MHD).  While this assumption is likely appropriate for many Be stars, for others there is evidence that disc feeding can suddenly turn on, leading to disc formation in B stars where there previously was none (e.g., \citealt{Guinan1984,Sonneborn1988}); suddenly turn off, leading to disc dissipation (e.g., \citealt{Wisniewski2010,Draper2011}); or be otherwise variable, with the relative strength of certain observational features that are linked to the presence of a disc growing or shrinking with time (e.g., \citealt{Hanuschik1993,Rivinius1998a}).  If the timescale for disc feeding variability is comparable to or smaller than the timescale for the disc to reach a steady state, then the dynamics of the disc will be constantly evolving in a complicated interplay between feeding, growth, and dissipation (e.g., \citealt{Haubois2012}).

We have shown results for only one particular set of pulsation parameters, listed in Table \ref{tab:parameters}. Moreover, some of the values of these parameter were particularly chosen to optimize disc formation; the rotation rate of the star is very close to critical ($W = 0.95$) 
and the magnitude of the velocity perturbation is high enough to reach critical rotation ($v_{\varphi,{\rm pert}} = 0.055 v_{\rm kep}[r =R_\star]$).  Evaluating whether or not the pulsational mass loss mechanism can operate efficiently and robustly for other parameters is beyond the scope of this work.  Instead, the main goals of this paper are 1) to show that non-radial pulsations can lead to quasi-steady viscous decretion disc at least under certain conditions, and 2) to study the dynamics of that disc.   Further work will be required in order to determine how our results precisely depend on stellar parameters.  

Our simulations are isothermal.  In reality, the temperature of the gas is determined by a complicated interplay between radiative cooling and irradiation by the star that depends on the precise density of the photosphere and the stellar spectrum.  Although some radiative thermodynamic modeling suggest that this process can be reasonably approximated by an isothermal gas with $T  = T_{\rm eff}/2$ (e.g., \citealt{Millar1998,Carciofi2008}), where $T_{\rm eff}$ is the effective temperature of the star, detailed modeling of spectra and local disc structure likely require non-isothermal calculations, especially at small radii \citep{McGill2011,McGill2013}.  

The temperature of the gas in our simulations is higher than expected for Be decretion discs.  This allows us to fully resolve both the disc itself and the MRI contained within the disc while evolving the simulation for several viscous times at the disc edge.   We expect that the results of the simulations will scale with temperature in the same way as 1D models (e.g., \citealt{Okazaki2001,Krticka2011}).  More precisely, a cooler temperature would result in a thinner disc and a larger disc radius ($r_{\rm disc} \propto T^{-1}$, Equation \ref{eq:r_disc}) with a smaller decretion rate ($\dot M$ $\propto$ $ T^{1/2}$), all else being equal.  Note that the decretion rates of Be stars are estimated to fall within the relatively wide range of $10^{-12}$--$10^{-9}$ $M_\odot$/yr \citep{Vieira2015}, so the values of $\approx 1.3 \times 10^{-10}$ $M_\odot$/yr and $6 \times 10^{-12}$ $M_\odot$/yr that we find in our MHD and hydrodynamics simulation, respectively, are reasonable even when extrapolated to more realistic, lower temperatures.  
Additionally, the higher temperature in our simulations also likely enhances the Reynolds stress to some degree by making spiral density waves less tightly wound in the radial direction (e.g. \citealt{Ju2017}).  This means that such waves are likely even less important for angular momentum transport than seen here. 

The reason that we used a relatively large temperature was to make the computational resource requirement manageable. The MHD simulation cost $\approx$ $9.2 \times 10^5$ SUs/core-hours on Pittsburgh Supercomputing Center's (PSC) Bridges-2 RM nodes (roughly equivalent to $5.1 \times 10^4$ SUs/node-hours on Texas Advanced Computing Center's (TACC) Stampede2 to reach $1.2 \times 10^4$ $t_\star$.   
The hydrodynamics simulations is significantly cheaper, since it has one less level of refinement, significantly higher time steps, and requires less work per cycle.  
Since the scale height of the disc scales as $\sim$ $T^{1/2}$ and the viscous time at the disc radius scales as $\sim$ $T^{-3/2} $ (Equations \ref{eq:r_disc} and \ref{eq:t_visc}), the cost of these simulations will scale as $T^{-3}$ in order to run for at least a couple of viscous times at the disc radius and to resolve the disc with as many cells in $\theta$ and the MRI with as many cells in both $\theta$ and $r$ as we do here (with the same number of cells in $\varphi$).  It is thus extremely expensive to reduce the temperature to more realistic values for Be stars.  For example, an MHD simulation with $T = 1.1 \times 10^4$ K (appropriate for $\mu$ Centauri-like parameters, see \S \ref{sec:units}) would cost a formidable $46$ billion SUs/core-hours on PSC's Bridges-2 (roughly $2.6$ billion core-hours on TACC's Stampede2). Obviously, doing so is not practical.  Future work will likely have to determine more precisely how the results depend on temperature by running simulations with both higher and lower values (though still much higher than what would be realistic) and then extrapolating the results using 1D models as a guide.  


\section{Discussion and Conclusions}
\label{sec:conc}

We have presented the results of 3D, isothermal hydrodynamic and MHD simulations of decretion discs fed by a pulsating inner boundary condition. Both simulations run long enough to reach states that are quasi-steady.  In the hydrodynamics simulation, this occurs through a repetitive cycle of mass being fed into the domain at small radii (a few $R_\star$),  then falling back onto the star during the process of circularization.  As a result, the decretion rate at large radii, $\dot M_{\rm out}$, is positive but small ($<$ 10 times the unimpeded mass input rate from the inner boundary), while decretion rate through the inner boundary, $\dot M_{\rm in}$, oscillates from large and positive to large and negative (Figure \ref{fig:mdot}).  
$\dot M_{\rm in}$ in the MHD simulation in contrast remains steady (and generally equal to the peak of the $\dot M_{\rm in}$ in the hydrodynamics simulation) and $\dot M_{\rm out}$ increases until the point that a balance of $\dot M_{\rm out} \approx \dot M_{\rm in}$ is reached.  Unlike the hydrodynamics simulation, where mass is concentrated within a few $R_\star$, mass is spread out more evenly in radius, forming an extended distribution (Figure \ref{fig:mhd_v_hydro_contour}).
The reason for this is that the material being fed into the domain by the pulsating boundary does not initially have enough angular momentum to reach large radii. 
Absent efficient angular momentum transport this leads to a disc confined to small radii that grows in mass with time until it becomes massive enough to interfere with the feeding process from the star.  
At that point gas begins to fall back onto the star until the disc drops enough mass for the feeding process to become efficient again, and the cycle continues.  
With efficient angular momentum transport, however, the disc can expand with radius and balance the input of mass from the boundary with an equally large decretion rate. 
In the hydrodynamics simulation, the Reynolds stress is very weak $\alpha_{\rm r} \lesssim 10^{-3}$, while in the MHD simulation the combination of Maxwell and Reynolds stress results in a total $\alpha \approx 0.06$ for $r \lesssim 30 R_\star$ (Figure \ref{fig:alpha_beta}).  Thus the MHD simulation is much more efficient at transporting angular momentum outwards than the hydrodynamics simulation, which allows the disc to expand in radius and not accumulate mass at small radii.

The Maxwell stress close to the midplane is approximately 3--4 times larger than the Reynolds stress in the MHD simulation, typical of MRI turbulence.
The significant Maxwell stress is caused by the MRI amplifying the initial magnetic field from $\beta_0 = 500$ to $\beta \approx 10$ near the surface of the star (Figure \ref{fig:alpha_beta}). 
Despite some indication of spiral structures in the disc (Figure \ref{fig:midplane_contour}), waves excited by the pulsating inner boundary condition are not able to coherently propagate outwards and thus do not result in a particularly large Reynolds stress.  
This conclusion is supported by a Fourier analysis of the temporal density variability, which reveals clear pattern speeds only at radii just outside of the star ($r \lesssim 4 R_\star$) with peaks at the harmonics of the stellar pulsation frequency; for larger radii the variability is not concentrated on individual frequencies.

Coherent spiral density waves have been invoked to explain long term V/R (i.e. violet/red) variability, where the two peaks in an emission line will vary in relative intensity on the time scale of $\gtrsim$ years \citep{Okazaki1991}.  The angular frequency of these waves is expected to be $\ll 0.1 (H/r)^2 \Omega$, corresponding to a period of $\gg$ $6300 t_\star$ for our parameters, with a similar growth time.  Over the course of the $\approx$ $1.2 \times 10^4 t_\star$ total run time of our MHD simulation, we do not see any evidence for the formation of a global, low frequency $m=1$ mode.
This may be because we do not include the appropriate excitation mechanism (e.g., radiation or further azimuthal asymmetry on the surface of the star), or because we neglect the effect of rotational deformation on the star.  The latter effect may be a key factor as it introduces a quadripolar component to the gravitational potential that could cause precession in the disc \citep{Papaloizou1992}. 

The 1D structure of the disc near the midplane in our MHD simulation is fairly well described by an isothermal viscous decretion disc model with an $\alpha$ parameter of 0.5 for $r \le r_{\rm c}$ and 0 for $r>r_{\rm c}$, where $r_{\rm c}$ is the critical radius (Figure \ref{fig:3D_1D_comp},  \citealt{Okazaki2001}).  Below the critical radius, $\approx$ 28--30 $R_\star$ for our parameters, we have a thin, subsonic, MRI turbulent, Keplerian disc that grows in thickness with radius.  Above the critical radius, we have a supersonic, angular momentum conserving, nearly spherical wind.  Significant departure from the 1D model occurs near the surface of the disc, however.  There, large-scale magnetic fields provide enough torque to accelerate a supersonic wind even at small radii (Figure \ref{fig:theta}).  This wind is strong enough to contribute significantly to the decretion rate despite the $\sim$ 10 times lower densities than the midplane, which is likely why the 1D model needs a larger $\alpha$ than what is measured in our simulations (Figure \ref{fig:alpha_beta}) in order to match the radial profiles.  

Actual temperatures (and thus, scale heights) in Be star discs are expected to be much smaller (by a factor of $>10$) than that assumed for our simulations.  Smaller gas temperature would push the radial transition from disc to wind out to much larger radii, estimated as $>1000 R_\star$ using Equation \ref{eq:r_disc} for the parameters of $\mu$ Centauri.   It is not as clear how the magnetically-torqued surface wind would be affected, though we do not expect the qualitative behavior to depend strongly on temperature.  If the MRI robustly amplifies the field to a value of $\beta$ roughly independent of temperature, then it is likely that the relative torque on the surface of the disc will remain unchanged and the angular structure we see in these simulations will be retained.    

This work serves as a proof-of-principle that the pulsationally driven orbital mass ejection model can lead to a Keplerian decretion disc as long as a weak magnetic field is present in the photosphere of the star.  Since the properties of this disc near the midplane generally agree with the 1D viscous decretion disc model, it possesses all the basic features needed to qualitatively explain the Be phenomenon. Nevertheless, the magnetically-torqued surface wind such as that which develops in our 3D simulations is not able to be captured in the 1D models and it is not clear how this could impact parameter estimation from observations.  For instance, our work suggests that a naive implementation of the 1D models would result in an over-estimation of the ``true'' $\alpha$ in the disc (Figure \ref{fig:alpha_beta}).   It may be thus be important for future observational modeling to consider this possibility.  It will also be important to evaluate the  fraction of Be stars for which the pulsationally driven orbital mass ejection model can apply and to more fully explore the parameter space of the pulsations and stellar magnetic field.

\section*{Acknowledgments}
I thank the anonymous referee for comments and suggestions that significantly improved the manuscript. 
I thank Lars Bildsten and Omer Blaes for useful discussions and comments on the manuscript. SMR was supported by the Gordon and Betty Moore Foundation through Grant GBMF7392.  SMR also thanks R. and D. Ressler for
their generous hospitality during the writing of this manuscript.
This research was supported in part by the National Science Foundation (NSF) under Grant No. NSF PHY-1748958, and by the NSF
through XSEDE computational time allocation TG--AST200005 on Stampede2.  This work was made possible by computing time granted by UCB on the Savio cluster.

\section*{Data Availability}
The data underlying this paper will be shared on reasonable request
to the corresponding author.

\bibliographystyle{mn2efix}
\bibliography{Be}

\appendix

\section{Grid}
\label{App:grid}

\begin{figure}
\includegraphics[width=0.45\textwidth]{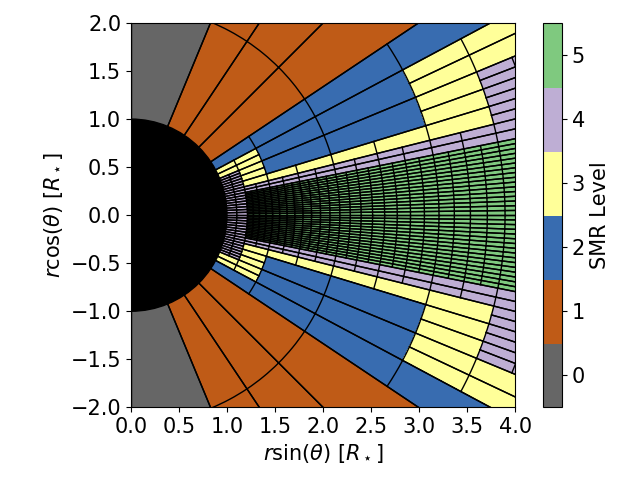}
\includegraphics[width=0.45\textwidth]{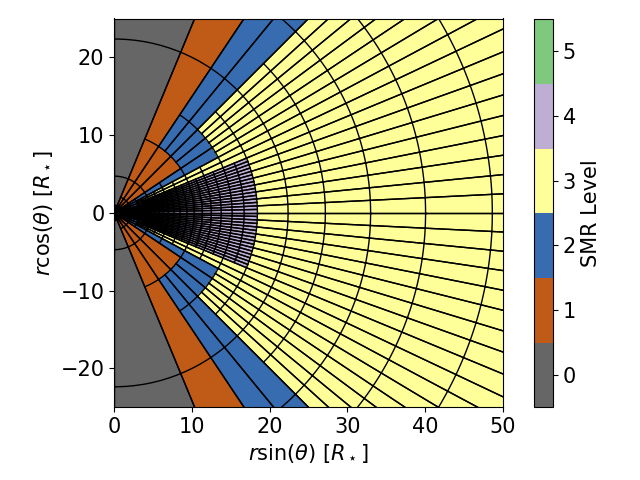}
\includegraphics[width=0.45\textwidth]{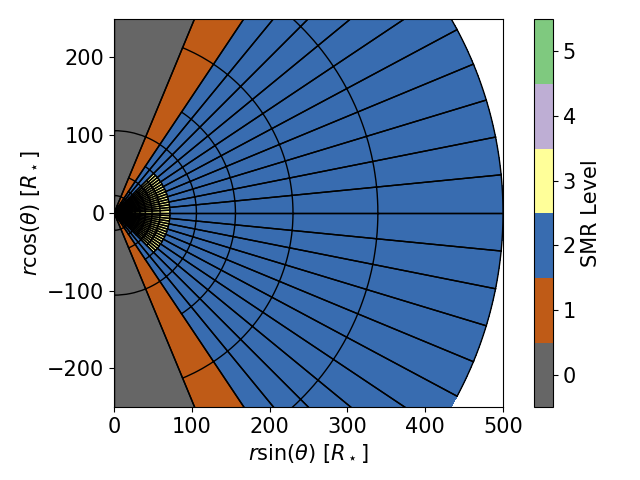}
\caption{SMR grid used in our MHD simulation, shown at small (top), medium (middle), and large (bottom) scale.  Colors represent the local level of refinement, with the $n$th level being $2^n$ times more resolved than the base $64 \times 32 \times 12$ resolution.  Each block represents $16 \times 4$ cells in $r$ and $\theta$, respectively.  The hydrodynamics simulation grid is similar, except that the highest level is 4.  This distribution of cells focuses resolution towards the high density midplane and away from the low density poles and keeps the number of cells per scale height approximately constant with radius. } 
\label{fig:grid}
\end{figure}

In this Appendix we describe our computational grid in more detail.  Table \ref{tab:resolution} lists the precise locations of each refinement level in both simulations, while Figure \ref{fig:grid} illustrates the grid on multiple scales for our MHD simulation.  The highest level of refinement is concentrated at small radii near the midplane with an angular extent such that it contains about 2 scale heights.  Every factor of $\sim$ 4 in radius we de-refine the midplane resolution by one level while increasing the width of the refinement region by a factor of 2 in polar angle.  This keeps the ratio between the local resolution and scale height approximately constant with radius.  We also ensure that there are no refinement boundaries in the region where mass is being inputted from the inner boundary at $r = R_\star$ by placing a level 4 refinement region between $3 {\rm \pi}/8 \le \theta \le 5 {\rm \pi}/8$.    The highest levels of refinement in the hydrodynamics and MHD simulations respectively are 4 and 5, where the higher resolution in the latter is used to ensure that the MRI is sufficiently resolved.  The hydrodynamics grid is similar to that shown in Figure \ref{fig:grid} except levels 2,3,4, and 5 are replaced by levels 1,2,3, and 4.  Where the MHD and hydrodynamics grids are equivalent in the regions of the MHD simulation covered by levels 0 and 1.  

\begin{table}
  \begin{center}
    \caption{Location of SMR levels in our simulations.  Between any two regions of refinement different by more than one level, every intervening level is achieved in transition.  For simplicity, we have not listed these transition regions that often have complicated shapes but they are shown in Figure \ref{fig:grid}.   }
    \label{tab:resolution}
    \def\arraystretch{1.75}
    \begin{tabular}{|c|c|c|c|} 
            \multicolumn{4}{c}{MHD} \\
            \hline
             Level & $r$-range & $\theta$-range  &  Effective Resolution\\
            \hline
                  $0$ & $R_\star \lesssim r \lesssim 500 R_\star$ & $0 \lesssim \theta \lesssim {\rm \pi}/8$ & $64 \times 32 \times 12$ \\
                   & $R_\star \lesssim r \lesssim 500 R_\star$ &  $ 7{\rm \pi}/8 \lesssim\theta \lesssim {\rm \pi}$&  \\
      \hline
       $1$ & $ R_\star \lesssim r \lesssim 500 R_\star$ & ${\rm \pi}/8 \lesssim \theta \lesssim 7{\rm \pi}/8$&$128 \times 256 \times 24$ \\
       \hline
       $2$ & $64 R_\star \lesssim r \lesssim 256 R_\star$& $3{\rm \pi}/16 \lesssim \theta \lesssim 13{\rm \pi}/16$& $256 \times 128 \times 48$\\
       \hline
        $3$ & $16 R_\star \lesssim r \lesssim 64 R_\star$  &${\rm \pi}/4\lesssim \theta \lesssim  3{\rm \pi}/4$  &$512 \times 256 \times 96$ \\
               \hline
        $4$ & $4 R_\star \lesssim r \lesssim 16 R_\star$& $3{\rm \pi}/8\lesssim \theta \lesssim 5{\rm \pi}/8$& $1024 \times 512 \times 192$\\
        & $R_\star \lesssim r \lesssim 1.2 R_\star$& $3{\rm \pi}/8 \lesssim  \theta \lesssim 5{\rm \pi}/8$&\\
        \hline
        $5$ & $1.2 R_\star \lesssim r \lesssim 4 R_\star$ & $7{\rm \pi}/16\lesssim \theta \lesssim 9{\rm \pi}/16$ &$2048 \times 1024 \times 384$ \\
        \hline
       \multicolumn{4}{c}{Hydro}\\
            \hline
            Level & $r$-range & $\theta$-range  &  Effective Resolution\\
            \hline
        $0$ & $R_\star \lesssim r \lesssim 500 R_\star$ & $0 \lesssim \theta \lesssim {\rm \pi}/8$ & $64 \times 32 \times 12$ \\
                   & $R_\star \lesssim r \lesssim 500 R_\star$ &  $ 7{\rm \pi}/8 \lesssim\theta \lesssim {\rm \pi}$&  \\
      \hline
       $1$ & $ R_\star \lesssim r \lesssim 500 R_\star$ & ${\rm \pi}/8 \lesssim \theta \lesssim 7{\rm \pi}/8$&$128 \times 256 \times 24$ \\
       & $64 R_\star \lesssim r \lesssim 256 R_\star$& $3{\rm \pi}/16 \lesssim \theta \lesssim 13{\rm \pi}/16$& \\
       \hline
        $2$ & $16 R_\star \lesssim r \lesssim 64 R_\star$  &${\rm \pi}/4\lesssim \theta \lesssim  3{\rm \pi}/4$  &$256 \times 128 \times 48$ \\
               \hline
        $3$ & $4 R_\star \lesssim r \lesssim 16 R_\star$& $3{\rm \pi}/8\lesssim \theta \lesssim 5{\rm \pi}/8$& $512 \times 256\times 96$\\
        \hline
        $4$ & $1.2 R_\star \lesssim r \lesssim 4 R_\star$ & $7{\rm \pi}/16\lesssim \theta \lesssim 9{\rm \pi}/16$ &$1024 \times 512 \times 192$ \\
        & $R_\star \lesssim r \lesssim 1.2 R_\star$& $3{\rm \pi}/8 \lesssim  \theta \lesssim 5{\rm \pi}/8$& \\
        \hline
    \end{tabular}
  \end{center}
\end{table}

\section{Density Floors and Mass Input}
\label{App:floor}
\begin{figure}
\includegraphics[width=0.45\textwidth]{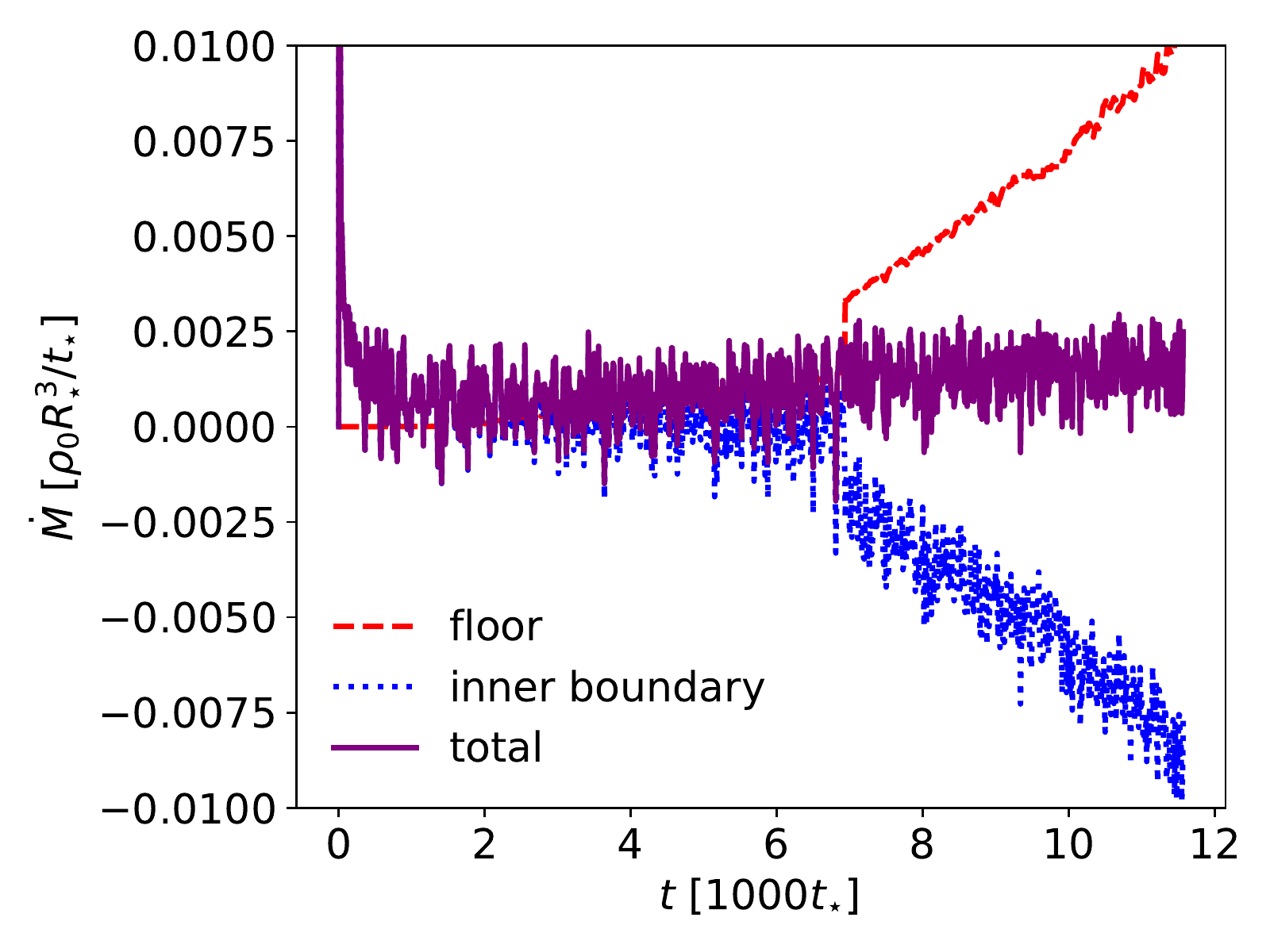}
\caption{Mass input rate from the floors (dashed), the inner boundary (dotted), and the sum of the two (solid) in our MHD simulation.  Initially, the floor input rate is negligible compared to the influx of mass from the inner boundary.  However, after $\sim$ 6000 $t_\star$, it becomes significant and grows approximately linearly with time.  Simultaneously, the input rate from the boundary becomes negative (i.e., mass is leaving the domain through the inner boundary) and decreases approximately linearly with time at the same rate as the floor input rate.  The sum of these two quantities, however, remains approximately constant with time.  This is because the floors produce mass at small radii in the low-density, highly magnetized polar regions, which is primarily just falling back onto the ``star'' (i.e., the inner boundary) as we highlight in Figure \ref{fig:floor_contours}.      } 
\label{fig:floor_v_t}
\end{figure}

In this Appendix we discuss the effects that numerical density floors have on our simulations and ensuing analysis.  In the hydrodynamics simulation, the net rate of mass added to the domain via the floors, $\dot M_{\rm floor}$, is negligible compared to the mass flux through the inner ($\dot M_{\rm in}$) and outer ($\dot M_{\rm out}$) boundaries.  
In contrast, $\dot M_{\rm floor}$ in the MHD simulation becomes substantial, where the floors depend on the local magnetic field strength that continuously grows with time in the polar regions\footnote{This occurs because the polar regions are inflowing, not outflowing, so that the field being added does not get carried away but builds up in strength.  If we were able to run the simulation for much longer times, presumably the magnetic pressure in the poles would eventually become comparable to the thermal pressure of the disc.  Whether the field then simply saturates and decretion proceeds just as before or if a new state of decretion would develop is unclear.  Investigating the latter possibility is reserved for future work.}, meaning that the density floors also continually grow in magnitude.  
This is seen in Figure \ref{fig:floor_v_t}, which plots $\dot M_{\rm floor}$ compared with the $\dot M_{\rm in}$.  At first the floor contribution is negligible, but eventually it becomes much larger than the initial $\dot M_{\rm in}$ and continues to grow with time.  At the same time, $\dot M_{\rm in}$  becomes large and negative, growing in magnitude at a rate that almost exactly mirrors the growth in $\dot M_{\rm floor}$.   As a consequence, $\dot M_{\rm in}+\dot M_{\rm floor}$ is $\approx$ constant with time, equal to the initial $\dot M_{\rm in}$ before floor contributions became significant.  

\begin{figure*}
\includegraphics[width=0.45\textwidth]{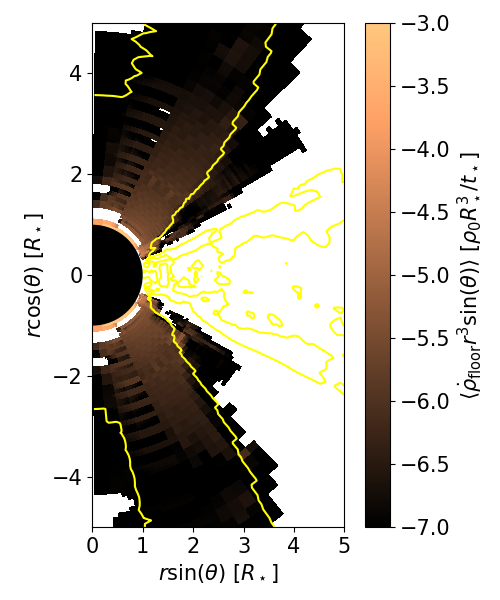}
\includegraphics[width=0.45\textwidth]{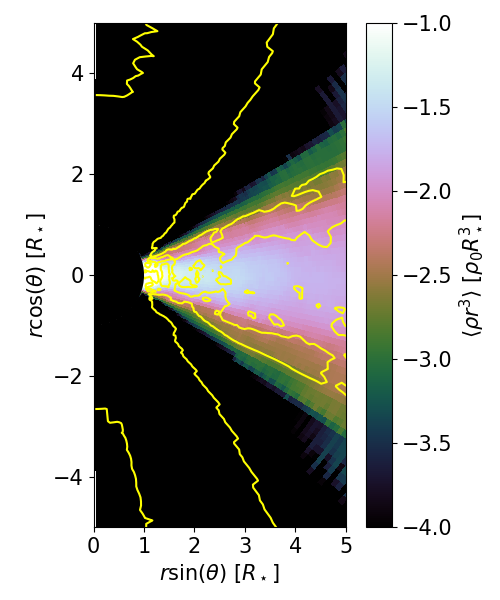}
\caption{Time and $\varphi$-averaged contours from our MHD simulation.  Left: rate at which the floors add mass density, $\langle \rho_{|rm floor} r^3 \sin(\theta)\rangle$.  Right: mass density, $\langle \rho r^3\rangle$.  Yellow lines trace contours of $\langle v_r \rangle=0$.   The majority of mass being added by the floors is located just outside $r = R_\star$ in regions that are dominated by inflow.  This means that most of this mass simply falls through the inner boundary.  Note that the regions near the midplane (where matter is flowing out of the inner boundary into the domain) are unaffected by the floors. } 
\label{fig:floor_contours}
\end{figure*}

To understand this behavior, Figure \ref{fig:floor_contours} shows a 2D contour of the azimuthal and time-averaged $\dot \rho_{\rm floor} r^3 \sin(\theta)$ over-plotted with contours of $v_r =0$ compared with a 2D contour of the averaged density.   The dominant contribution to $\dot M_{\rm floor}$ happens just outside the inner boundary in the polar regions, away from the mass of the disc where the flow is predominantly inflowing.  Thus, the mass getting added by the floor simply flows out of the domain through the polar region of the inner radial boundary without affecting the outflowing disc or surface wind.   The calculation of the net $\dot M_{\rm in}$ is then obscured so that the more meaningful quantity is $\dot M_{\rm in} + \dot M_{\rm floor}$, which represents the amount of mass being added by the boundary independently of the floors.  This is what is used in the main text (e.g., in Figure \ref{fig:mdot}).


\end{document}